\documentclass[12pt,paper=a4,DIV=15]{scrartcl}
\pdfoutput=1
\usepackage[british]{babel}
\usepackage{lmodern}
\usepackage{pstricks}
\usepackage{xcolor}
\definecolor{darkblue}{rgb}{0,0,0.5}
\definecolor{Pblue}{rgb}{0.37,0.51,0.71}
\definecolor{Pred}{rgb}{0.92,0.39,0.21}
\definecolor{Pyellow}{rgb}{0.88,0.61,0.14}
\definecolor{Pgreen}{rgb}{0.56,0.69,0.19}
\usepackage[colorlinks,linkcolor=darkblue]{hyperref}
\usepackage{amssymb,amsmath,bm,bbm}
\usepackage{epsf}
\usepackage{epsfig}
\usepackage{afterpage}
\usepackage{longtable}
\usepackage{cite}
\usepackage{latexsym,mathrsfs,dsfont}
\usepackage{graphics}
\usepackage{url}
\usepackage{paralist}
\usepackage{bbold}
\usepackage{enumerate}
\usepackage{slashed}
\usepackage{hyphenat}
\addtokomafont{caption}{\small}

\DeclareMathAlphabet{\mathsfit}{\encodingdefault}{\sfdefault}{m}{sl}
\SetMathAlphabet{\mathsfit}{bold}{\encodingdefault}{\sfdefault}{bx}{sl}

\newcommand{\bldm}[1]{\bm{\mathsfit{#1}}}

\newcommand{\gev}{\, \text{GeV}}

\newcommand{\beq}{\begin{equation}}
\newcommand{\eeq}{\end{equation}}
\newcommand{\be}{\begin{equation}}
\newcommand{\ee}{\end{equation}}
\newcommand{\bi}{\begin{itemize}}
\newcommand{\ei}{\end{itemize}}
\newcommand{\ba}{\begin{array}}
\newcommand{\ea}{\end{array}}
\newcommand{\beqa}{\begin{eqnarray}}
\newcommand{\eeqa}{\end{eqnarray}}
\newcommand{\bea}{\begin{eqnarray}}
\newcommand{\eea}{\end{eqnarray}}
\newcommand{\beqn}{\begin{eqnarray}}
\newcommand{\eeqn}{\end{eqnarray}}

\def\kpn{K^+\rightarrow\pi^+\nu\bar\nu}
\def\klpn{K_{L}\rightarrow\pi^0\nu\bar\nu}

\begin{document}

\begin{flushright}
{\small FLAVOUR(267104)-ERC-80}
\end{flushright}

\medskip

\begin{center}
{\Large\bfseries\sffamily
$\bldm{B\to K^{(*)}\nu\bar\nu}$ decays in the Standard Model and beyond}
\\[0.8 cm]
{\sffamily Andrzej~J.~Buras\,$^{a,b}$, Jennifer Girrbach-Noe\,$^{a,b}$, Christoph Niehoff\,$^{c}$,
 and  David M. Straub\,$^{c}$
 \\[0.5 cm]}
{\footnotesize\slshape
$^a$TUM Institute for Advanced Study, Lichtenbergstr. 2a, 85748 Garching, Germany\\
$^b$Physik Department, TUM,
James-Franck-Stra{\ss}e, 85748 Garching, Germany\\
$^c$Excellence Cluster Universe, TUM,
Boltzmannstr.~2, 85748~Garching, Germany}
\vskip0.41cm

{\footnotesize
E-Mail:
\texttt{\href{mailto:andrzej.buras@tum.de}{andrzej.buras@tum.de}},
\texttt{\href{mailto:jennifer.girrbach@tum.de}{jennifer.girrbach@tum.de}},
\texttt{\href{mailto:christoph.niehoff@tum.de}{christoph.niehoff@tum.de}},
\texttt{\href{mailto:david.straub@tum.de}{david.straub@tum.de}}
}
\end{center}
\smallskip

\setcounter{tocdepth}{2}

\abstract{\noindent
We present an analysis of the rare exclusive $B$ decays $B\to K\nu\bar\nu$ and $B\to K^{*}\nu\bar\nu$
within the Standard Model (SM), in a model-independent manner, and in a number of new physics (NP) models.
Combining new form factor determinations from lattice QCD with light-cone sum rule results and including complete two-loop electroweak corrections to the SM Wilson coefficient, we obtain the SM predictions
$\text{BR}(B^+\to K^+\nu\bar\nu) = (4.0 \pm 0.5) \times 10^{-6}$
and
$\text{BR}(B^0\to K^{* 0}\nu\bar\nu)  =  (9.2\pm1.0) \times 10^{-6}$,
more precise and more robust than previous estimates.
Beyond the SM, we make use of an effective theory with dimension-six operators invariant under the SM gauge symmetries to relate NP effects in $b\to s\nu\bar\nu$ transitions to $b\to s\ell^+\ell^-$ transitions and use the wealth of experimental data on $B\to K^{(*)}\ell^+\ell^-$ and related modes to constrain
NP effects in $B\to K^{(*)}\nu\bar\nu$.
We then consider several specific NP models, including $Z'$ models, the MSSM, models with partial compositeness, and leptoquark models, 
demonstrating that 
 the correlations between $b\to s\nu\bar\nu$ 
observables among themselves and with 
$B_s\to\mu^+\mu^-$ and 
$b\to s\ell^+\ell^-$ transitions offer 
powerful tests of NP with new right-handed couplings and non-MFV interactions.
}

\thispagestyle{empty}

\tableofcontents

\section{Introduction}

Rare $K$ and $B$ decays with a neutrino pair in the final state, belonging to 
the theoretically cleanest in the field of flavour-changing neutral current (FCNC) processes, should soon play
an important role in the tests of the Standard Model (SM) and its extensions\footnote{Recent reviews can be found in \cite{Buras:2013ooa,Isidori:2014rba}.}
This is due to a number of experiments being either planned or in 
preparation. Provided the rates of these decays are not significantly smaller 
than predicted within the SM, at the end of this decade 
we should have rather precise measurements of their branching ratios at our disposal.

As the search for new physics (NP) through these decays is based on possible deviations from SM predictions, it is crucial that the latter are as precise as possible.
In the case of the decays $\kpn$ and $\klpn$, the hadronic uncertainties 
are very small as the relevant hadronic matrix elements can be extracted from the leading semi-leptonic $K^+$ and $K_L$ decays using isospin symmetry. For the decays $B\to K^*\nu\bar\nu$ and 
$B\to K \nu\bar\nu$, this is not possible, so studying them requires the evaluation of the relevant form factors by means of non-perturbative methods. The corresponding perturbative, short distance QCD and electroweak effects are also important but as we will summarize below they are by now fully under control.
Most importantly, the decays based on the $b\to s\nu\bar\nu$ transition do not suffer from hadronic uncertainties {\em beyond} the form factors, that plague the $b\to s\ell^+\ell^-$ transitions due to the breaking of factorization caused by photon exchange. For the $B\to K^{(*)}\nu\bar\nu$ transitions, factorization is exact, so a measurement of the decay rates would allow in principle to measure the form factors.
In the last two years, lattice computations of $B\to K$ and $B\to K^*$ form factors have become available \cite{Bouchard:2013eph,Horgan:2013hoa} that are valid at large $q^2$ of the neutrino pair and which complement the existing results within light-cone sum rules (LCSR) \cite{Ball:2004ye,Ball:2004rg,Khodjamirian:2010vf}, valid at low and intermediate $q^2$. Combining these two sources of information, we will give SM predictions for the observables valid in the entire kinematic range, not relying on model-dependent extrapolations.

The relation between $b\to s\nu\bar\nu$ and $b \to s\ell^+\ell^-$ processes is not just relevant in the SM, where they are governed by 
the same form factors, but also beyond the SM, since the $SU(2)_L$ gauge symmetry relates neutrinos to left-handed charged leptons. The absence of any
direct NP signal close to the electroweak scale at the LHC implies that this symmetry should still be reflected approximately in low-energy observables.
 This fact can be exploited by considering dimension-6 operators of SM fields invariant under the SM gauge symmetry\footnote{Such EFT approach has received increasing interest in the context of flavour physics 
recently, see e.g.~\cite{Crivellin:2014zpa,Alonso:2014csa,Hiller:2014yaa}.}, 
some of which contribute to $b\to s\nu\bar\nu$ 
and $b \to s\ell^+\ell^-$ processes simultaneously. 
This correlation
is particularly interesting in view of various tensions with the SM recently observed in exclusive $b \to s\ell^+\ell^-$ decays \cite{Aaij:2013qta,Descotes-Genon:2013wba,Altmannshofer:2013foa,Beaujean:2013soa,Lyon:2014hpa,Altmannshofer:2014rta}
that, if due to NP, might also leave an imprint in $b\to s\nu\bar\nu$ decays.
Although on a completely model-independent basis, no general conclusions can be drawn,
it turns out that in specific NP models,
often only a subset of these operators are present and we will demonstrate for several models that 
clear-cut predictions for the size of the effects and various correlations can be obtained.

In 2009, a detailed analysis of $B\to K^*\nu\bar\nu$, $B\to K \nu\bar\nu$ and $B\to X_s\nu\bar\nu$ has been presented \cite{Altmannshofer:2009ma}, giving the SM predictions and studying correlations among these decays as well as with $s \to d\nu\bar\nu$ and $b\to s\ell^+\ell^-$ processes.
Most importantly, it has been pointed out that  these correlations offer 
powerful tests of NP with new right-handed couplings and non-MFV interactions.
Several of the formulae presented in that paper have been used since then
in the study of a number of NP models
\cite{Buras:2010pz,Buras:2012jb,Buras:2013qja,Buras:2012dp,Straub:2013zca,Buras:2013dea,Buras:2014yna,Biancofiore:2014uba}. For earlier studies of $b\to s\nu\bar\nu$ transitions, see 
in particular \cite{Colangelo:1996ay,Buchalla:2000sk}.

As the flavour precision era 
will certainly be one of the frontiers of particle physics in the second 
half of this decade, it is the right time to have a closer look at these 
decays with the goal to improve the accuracy of SM predictions and to 
generalize the NP study beyond the one presented in \cite{Altmannshofer:2009ma}. Here we summarize the main novelties in our present paper.
\begin{itemize}
\item We update the SM predictions for the $B\to K^{(*)}\nu\bar\nu$ branching ratios and the angular observable $F_L$ in $B\to K^*\nu\bar\nu$, using a combined fit of the $B\to K$ and $B\to K^*$ form factors to LCSR and lattice calculations, making the SM predictions not only more precise, but also more reliable;
\item We include the  complete two-loop electroweak corrections to the SM Wilson coefficient;
\item We discuss the model-independent implications of the precise measurements of $B\to K^*\mu^+\mu^-$ and $B\to K\mu^+\mu^-$ observables at LHCb on $B\to K^{(*)}\nu\bar\nu$, using dimension-6 operators invariant under the SM gauge symmetry;
\item We investigate the impact of departing from lepton flavour universality;
\item We discuss predictions for  $B\to K^{(*)}\nu\bar\nu$ in several NP models, including $Z'$ models, the MSSM, and models with scalar or vector leptoquarks;
\item We stress and demonstrate both in the model-independent approach and in the context of general $Z^\prime$ models that the correlations of the recently measured rate 
for $B_s\to\mu^+\mu^-$ with the rates for  $B\to K^{(*)}\nu\bar\nu$ and  $B\to K^{(*)}\mu^+\mu^-$ offer particularly powerful tests of the presence of right-handed currents. 
  \item
 We point out that the decays $B\to K^{(*)}\nu\bar\nu$ allow to distinguish between $Z$ and $Z^\prime$ explanations of the present departures 
 of the data from SM predictions for  $B_s\to\mu^+\mu^-$, $B\to K^*\mu^+\mu^-$ and $B\to K\mu^+\mu^-$ that is rather hard on the basis of these three decays alone.
 \end{itemize}

Our paper is organized as follows. In section~\ref{sec:2} we define most 
important observables and present improved results for them within the SM. 
In section~\ref{sec:3} we present a general discussion of NP in the framework 
of effective theories. In  section~\ref{sec:4} we will present the results 
in a number of concrete NP models stressing the importance of correlations 
between various observables that  allow to distinguish between these models. 
We summarize the main results of our paper and conclude in section~\ref{sec:5}.
 The detailed information on form factors and various definitions are 
relegated to appendix~\ref{sec:FF}.
\section{Standard Model}\label{sec:2}

\subsection{Effective Hamiltonian and observables}

The effective Hamiltonian for $b\to s\nu\bar\nu$ transitions in the SM reads
\begin{equation} \label{eq:Heff}
{\mathcal{H}}_{\text{eff}}^\text{SM} = - \frac{4\,G_F}{\sqrt{2}}V_{tb}V_{ts}^*
C_L^\text{SM} \mathcal O_L
~+~ \text{h.c.} \,,
\end{equation}
where%
\begin{align}
\mathcal{O}_{L} &=\frac{e^2}{16\pi^2}
(\bar{s}  \gamma_{\mu} P_L b)(  \bar{\nu} \gamma^\mu(1- \gamma_5) \nu)
\,.
\end{align}
The Wilson coefficient $C_L^\text{SM}$ is known with a high accuracy, including NLO QCD corrections \cite{Buchalla:1993bv,Misiak:1999yg,Buchalla:1998ba} and two-loop electroweak contributions \cite{Brod:2010hi}, resulting in
\begin{align}
C_L^\text{SM} &= -X_t/s_w^2
\,,&
X_t&=1.469\pm0.017
\,.
\label{eq:CLSM}
\end{align}

Since the neutrinos escape the detector unmeasured, there are three observables that can be measured in the decays $B\to K^{(*)}\nu\bar\nu$ as functions of $q^2$: the two differential branching ratios and the $K^*$ longitudinal polarization fraction $F_L$ in $B\to K^{*}\nu\bar\nu$, first suggested in \cite{Altmannshofer:2009ma}. In the SM, they can be written as
\begin{align}
\frac{d\text{BR}(B^+\to K^+\nu\bar\nu)_\text{SM}}{dq^2}
&\equiv \mathcal{B}_{K}^\text{SM}(q^2)
= \tau_{B^+} 3|N|^2\frac{X_t^2}{s_w^4} \rho_{K}(q^2),
\\
\frac{d\text{BR}(B^0\to K^{*0}\nu\bar\nu)_\text{SM}}{dq^2}
&\equiv \mathcal{B}_{K^*}^\text{SM}(q^2)
= \tau_{B^0}3|N|^2\frac{X_t^2}{s_w^4}  \left[\rho_{A_{1}}(q^2)+\rho_{A_{12}}(q^2)+\rho_V(q^2)\right],
\\
F_L(B \to K^*\nu\bar\nu)_\text{SM}
&\equiv F_L^\text{SM}(q^2)
= \frac{\rho_{A_{12}}(q^2)}{\rho_{A_{1}}(q^2)+\rho_{A_{12}}(q^2)+\rho_V(q^2)},
\label{eq:FLSM}
\end{align}
where the factor of 3 stems from the sum over neutrino flavours,
\begin{equation}
N= V_{tb}V_{ts}^*\,
\frac{G_F \alpha}{16\pi^2 }
\sqrt{\frac{m_B}{3\pi}}\,,
\end{equation}
is a normalization factor and the $\rho_i$ are rescaled form factors defined in appendix~\ref{sec:kappa}.
In contrast to $B\to K^{(*)}\ell^+\ell^-$ decays, the isospin asymmetries of the decays with neutrinos in the final state  vanish identically, so the branching ratio of the $B^0$ and $B^\pm$ decays only differ due to the lifetime difference. $F_L$ is equal for charged and neutral $B$ decay.

We also define $q^2$-binned observables
\begin{align}
\left\langle\mathcal{B}_{K^{(*)}}^\text{SM}\right\rangle_{[a,b]}
&\equiv
\int_a^b dq^2 ~\frac{d\text{BR}(B\to K^{(*)}\nu\bar\nu)_\text{SM}}{dq^2}
\,,
\\
\left\langle F_L^\text{SM}\right\rangle_{[a,b]}
&\equiv
\frac{\int_a^b dq^2 \rho_{A_{12}}(q^2)}{\int_a^b dq^2 \left[\rho_{A_{1}}(q^2)+\rho_{A_{12}}(q^2)+\rho_V(q^2)\right]}
\,.
\end{align}

\subsection{Numerical analysis}

The numerical prediction of the observables within the SM requires the calculation of the hadronic form factors. Both for $B\to K$ and $B\to K^*$, lattice computations have become available recently \cite{Bouchard:2013eph,Horgan:2013hoa}, that are valid at large $q^2$. At low $q^2$, we make use of the results from light-cone sum rules \cite{Ball:2004ye,Ball:2004rg}.
Since the form factors have to be smooth functions of $q^2$, one can obtain expressions valid in the whole kinematical range relevant for $B\to K^{(*)}\nu\bar\nu$ by performing a combined fit to lattice and LCSR results. Since this approach makes use of theoretical input on both ends of the kinematical range, the results will be very weakly dependent on the parametrization chosen for the form factors. In the case of $B\to K^*$, such combined fit has been performed recently in \cite{BSZ}. For $B\to K$, we have performed our own fit that we discuss in detail in appendix~\ref{sec:FF}.

\begin{table}
\renewcommand{\arraystretch}{1.3}
\centering
\begin{tabular}{llllll}
\hline
$s_w^2$ & $0.23126(5)$ & \cite{Agashe:2014kda,Brod:2010hi} & $\tau_{B^0}$ &$1.519(5)$ ps &\cite{Agashe:2014kda}

\\
$\alpha$ & $127.925(16)$ & \cite{Agashe:2014kda,Brod:2010hi} & $\tau_{B^+}$ &$1.638(4)$ ps &\cite{Agashe:2014kda}
\\
$|V_{tb}V_{ts}^*|$ &
$0.0401(10)$ & \cite{Bona:2006ah,UTfit}
& $m_b^\text{1S}$&$4.66(3)$ GeV  &\cite{Agashe:2014kda}
\\
\hline
\end{tabular}
\caption{Input parameters used for the SM predictions.}
\label{tab:input}
\end{table}

Our numerical results for the differential branching ratios and $F_L$ are given in table~\ref{tab:kappa} for different bins of $q^2$ and for the whole kinematical region. The total branching ratios in the SM are shown in the last row, as the results of integrating over the whole kinematically allowed region,
\begin{align}
\text{BR}(B^+\to K^+\nu\bar\nu)_\text{SM} & =  (3.98 \pm 0.43 \pm 0.19) \times 10^{-6}, \\
\text{BR}(B^0\to K^{* 0}\nu\bar\nu)_\text{SM} & =  (9.19\pm0.86\pm0.50) \times 10^{-6}, \\
F^\text{SM}_L &= 0.47 \pm 0.03\,,
\end{align}
where the first error is due to the form factors and the second one parametric. From  table~\ref{tab:input},  summarizing  the numerical input used for these predictions, one can see that the parametric error is completely dominated by CKM elements.
We note that, when determined from tree-level decays, the uncertainty on the CKM combination $|V_{tb}V_{ts}^*|$ is dominated by the uncertainty on $|V_{cb}|$ and our value corresponds to $|V_{cb}|=0.0409(10)$. Using instead the PDG averages of the inclusive or exlusive determinations, respectively, which are at a $2.5\sigma$ tension with each other \cite{Agashe:2014kda}, the central values of the branching ratios would shift by 7\% up or down, respectively \cite{Altmannshofer:2014rta}.

As seen from (\ref{eq:FLSM}), in $F_L$ all these parametric uncertainties cancel, so we only quote the form factor uncertainty.
Note that the value of $F_L$ at the kinematical endpoints is fixed \cite{Altmannshofer:2009ma,Hiller:2013cza}.

\begin{table}[tbp]
\renewcommand{\arraystretch}{1.2}
\centering
\begin{tabular}{ccccc}
\hline
$q^2$ [GeV]$^2$ & $10^6\left\langle\mathcal{B}_{K^{*}}^\text{SM}\right\rangle$ & $\kappa_{\eta}$ & $\left\langle F_L^\text{SM}\right\rangle$ & $10^6\left\langle\mathcal{B}_{K}^\text{SM}\right\rangle$\\
\hline
$0-4$ &
$1.38\pm0.21\pm0.07$ & $1.64\pm0.04$ & $0.79\pm0.03$
& $0.93 \pm 0.14 \pm 0.05$
\\
$4-8$ & 
$1.88\pm0.22\pm0.10$ & $1.28\pm0.05$ & $0.56\pm0.03$
& $0.92 \pm 0.11 \pm 0.04$
\\
$8-12$ & 
$2.27\pm0.22\pm0.12$ & $1.16\pm0.05$ & $0.43\pm0.03$
& $0.86 \pm 0.09 \pm 0.04$
\\
$12-16$ &  
$2.36\pm0.18\pm0.13$ & $1.24\pm0.05$ & $0.35\pm0.02$
& $0.71 \pm 0.07 \pm 0.03$
\\
$16-q^2_\text{max}$ &
$1.30\pm0.10\pm0.07$ & $1.57\pm0.05$ & $0.32\pm0.03$
& $0.55 \pm 0.05 \pm 0.04$
\\
\hline
$0-q^2_\text{max}$ &
$9.19\pm0.86\pm0.50$ & $1.34\pm0.04$ & $0.47\pm0.03$
& $3.98 \pm 0.43 \pm 0.19$
\\
\hline
\end{tabular}
\caption{Results for various quantities as defined in the text. $q^2_\text{max}$ is the kinematic limit of 22.9~GeV in the case of $B\to K\nu\bar\nu$ and 19.2~GeV for $B\to K^*\nu\bar\nu$. For the differential branching ratios, the first error is due to the form factors and the second one due to parametric uncertainties (including the parameters in table~\ref{tab:input} as well as $X_t$). For the other quantities, parametric uncertainties are negligible.}
\label{tab:kappa}
\end{table}

Our result for the $B^+\to K^+\nu\bar\nu$ branching ratio is compatible with -- but significantly more precise 
than -- earlier determinations \cite{Altmannshofer:2009ma,Bartsch:2009qp}. In the case of the $B\to K^{*}\nu\bar\nu$ branching ratio, our new prediction is 
roughly 35\% higher than the one of \cite{Altmannshofer:2009ma}. There are two main sources for this difference. First, we use the electromagnetic fine 
structure constant $\alpha_\text{em}$ in the $\overline{\text{MS}}$ scheme at the scale $M_Z$ rather than at zero momentum transfer. This is the correct choice to be 
used with the Wilson coefficient including NLO electroweak corrections in (\ref{eq:CLSM}). In 2009, these corrections were not known yet, so the scheme and scale choice of $\alpha_\text{em}$ was a higher order effect. Second, the normalization of our form factors is fixed by the predictions of LCSR combined with the lattice predictions (which are in very good agreement). In \cite{Altmannshofer:2009ma}, the LCSR prediction for the form 
factors was only used for the form factor shape and the relative normalization, 
while the overall normalization was extracted from the experimental measurement of $\text{BR}(B\to K^*\gamma)$, following \cite{Ball:2006eu,Altmannshofer:2008dz}. 
We do not follow this approach because it assumes the absence of NP in $B\to K^*\gamma$. We also note that up-to-date experimental and 
theoretical predictions for $B\to K^*\gamma$ are in good agreement \cite{BSZ}.
Finally, we note that the error estimates of \cite{Altmannshofer:2009ma} did not include the uncertainty due to the model dependence of 
the extrapolation of the LCSR form factors to high $q^2$. Our new predictions include this uncertainty (which is much reduced by the addition of the lattice data) and
so they are not only more precise, but are  also put on a more firm footing.

Our predictions should be compared with the present experimental upper bounds.
Combining two analyses with hadronic or semi-leptonic tagging as well as charged and neutral $B$ decays, the BaBar collaboration finds \cite{Lees:2013kla}
\begin{align}
\text{BR}(B^+ \to K^+\nu\bar\nu) &<1.7\times 10^{-5}\ \text{(90\%\, CL)}.
\end{align}
The strongest bound on the $B\to K^*\nu\bar\nu$ decay was set by the Belle collaboration,
\cite{Lutz:2013ftz}
\begin{align}
\text{BR}(B^0 \to K^{*0}\nu\bar\nu) &<5.5\times 10^{-5}\ \text{(90\%\, CL)},
\\
\text{BR}(B^+ \to K^{*+}\nu\bar\nu) &<4.0\times 10^{-5}\ \text{(90\%\, CL)}.
\end{align}
Since $\text{BR}(B^+ \to K^{(*)+}\nu\bar\nu)/\text{BR}(B^0 \to K^{(*)0}\nu\bar\nu)=\tau_{B^+}/\tau_{B^0}$ holds in the SM and beyond, we can use the stronger of these bounds and obtain
\begin{align}
\mathcal R_K  \equiv \frac{\mathcal{B}_K}{\mathcal{B}_K^\text{SM}}& < 4.3\,,
&
\mathcal R_{K^*}\equiv \frac{\mathcal{B}_{K^*}}{\mathcal{B}_{K^*}^\text{SM}}  & < 4.4\,,
\end{align}
at 90\% C.L., where we have neglected the theory uncertainty.

At the Belle-II experiment, a conservative estimate of the sensitivity with 50~ab$^{-1}$, that is expected to be collected by 2023, envisages a measurement of 
the SM branching ratios with 30\% precision \cite{Aushev:2010bq} based on the predictions of \cite{Altmannshofer:2009ma}. Since we predict a significantly higher branching 
ratio for $B\to K^{*}\nu\bar\nu$ as discussed above, a better relative precision could be reached in turn. Moreover, already with 20~ab$^{-1}$, that is expected to 
be collected by 2020, 
first signs of NP could in principle be seen. Indeed, we will see in sections~\ref{sec:3} and \ref{sec:4} that in several NP models, the experimental upper bounds
can be saturated. In this case, a $5\sigma$ discovery should be definitely possible at Belle-II.

\section{Model-independent new physics analysis}\label{sec:3}

\subsection{Low-energy effective theory}

Beyond the SM (but assuming no NP lighter than the $B$ meson), a second operator can appear in the effective low-energy Hamiltonian for $b\to s\nu\bar\nu$ transitions,
\begin{equation} \label{eq:Heff1}
{\mathcal{H}}_{\text{eff}} = - \frac{4\,G_F}{\sqrt{2}}V_{tb}V_{ts}^*\left(C_L \mathcal O_L +C_R \mathcal O_R  \right) ~+~ \text{h.c.} \,,
\end{equation}
where
\begin{align}
\mathcal{O}_{L} &=\frac{e^2}{16\pi^2}
(\bar{s}  \gamma_{\mu} P_L b)(  \bar{\nu} \gamma^\mu(1- \gamma_5) \nu)
\,,&
\mathcal{O}_{R} &=\frac{e^2}{16\pi^2}(\bar{s}  \gamma_{\mu} P_R b)(  \bar{\nu} \gamma^\mu(1- \gamma_5) \nu)
\,.
\end{align}
In writing this effective Hamiltonian, we have explicitly assumed lepton flavour universality (LFU), i.e. that NP couples to all three neutrino flavours in the same manner. The implications of relaxing this assumption
will be discussed in general terms at the end of  this section and in section~\ref{sec:4} in the context of  leptoquark models.

In spite of the presence of two complex Wilson coefficients, the modification of the three observables can be described in terms of two real quantities $\epsilon>0$ and $\eta\in[-\frac{1}{2},\frac{1}{2}]$, defined as
\begin{equation}  \label{eq:epsetadef}
 \epsilon = \frac{\sqrt{ |C_L|^2 + |C_R|^2}}{|C_L^\text{SM}|}~, \qquad
 \eta = \frac{-\text{Re}\left(C_L C_R^{*}\right)}{|C_L|^2 + |C_R|^2}~,
\end{equation}
such that $\epsilon=1$ in the SM and $\eta\neq 0$ signals the presence of right-handed currents.
One finds
\begin{align}
 \mathcal{R}_K   & = (1 - 2\,\eta)\epsilon^2
 \,, &
 \mathcal{R}_{K^*} 
  & =
  (1 +  \kappa_\eta \eta)\epsilon^2
  \,, &
 \mathcal{R}_{F_L} \equiv \frac{F_L}{F_L^\text{SM}} 
 & =  
  \frac{1+2\eta}{1+\kappa_\eta\eta}
 \,.
\label{eq:epseta-R}
\end{align}
The parameter $\kappa_\eta$ depends on the form factors and its explicit form is given in appendix~\ref{sec:kappa}.
Its numerical value is listed in the last row of table~\ref{tab:kappa}.

Since the three observables in (\ref{eq:epseta-R}) only depend on two combinations of Wilson coefficients, there is a model-independent prediction,
\begin{equation}
F_L = F_L^\text{SM}
\left(\frac{(\kappa_\eta-2)\mathcal{R}_K+4\,\mathcal{R}_{K^*}}{(\kappa_\eta+2)\mathcal{R}_{K^*}}\right)
\,.
\label{eq:FLtest}
\end{equation}
In principle, this relation can be tested experimentally (also on a bin-by-bin basis). A similar relation can be obtained for the modification of the inclusive $B\to X_s\nu\bar\nu$ branching ratio,
\begin{equation}
\text{BR}(B\to X_s\nu\bar\nu)
\approx
\text{BR}(B\to X_s\nu\bar\nu)_\text{SM}\left(
\frac{\kappa_\eta \mathcal{R}_K+2\,\mathcal{R}_{K^*}}{\kappa_\eta+2}
\right)\,,
\label{eq:Xstest}
\end{equation}
where we have neglected a contribution proportional to $\eta$ of at most $\pm5\%$ to the inclusive branching ratio \cite{Altmannshofer:2009ma}.
Following \cite{Altmannshofer:2009ma} and using our updated numerical input, we obtain
\begin{equation}
\text{BR}(B\to X_s\nu\bar\nu)_\text{SM}
= 
(2.9\pm0.3)\times 10^{-5} \,.
\end{equation}
In section~\ref{sec:lfu}, we will show that the relations (\ref{eq:FLtest}) and (\ref{eq:Xstest}) hold even in the case of lepton flavour non-universality and lepton flavour violation. Consequently, a violation of either of them unambiguously signals the presence of particles other than neutrinos in the final state (as discussed e.g.\ in \cite{Altmannshofer:2009ma,Schmidt-Hoberg:2013hba}).

\subsection{Standard Model gauge-invariant effective theory}\label{sec:smeft}

As mentioned in the introduction, the $b\to s\nu\bar\nu$ transition is closely related to the $b\to s\ell^+\ell^-$ transition, on which there is a wealth of experimental data from exclusive and inclusive $B$ decays.
The reason for this correlation is that the neutrinos and left-handed charged leptons are related by $SU(2)_L$ symmetry.
To study these correlations in a model-independent manner, one can consider an operator product expansion with dimension-six operators invariant under 
the full SM gauge symmetry. This corresponds to an effective theory where all the SM degrees of freedom are kept as dynamical degrees of freedom and only the NP is integrated out, and we will refer to this EFT as SM-EFT in the following. This approach is meaningful if there is a separation of scales between the electroweak scale $v$ and the NP scale $\Lambda$, as is suggested by the absence of any new particles close to the electroweak scale in LHC searches so far. In fact, even when there are relatively light new particles, such as few-hundred GeV neutralinos and charginos in the MSSM, this effective theory turns out to be well-behaved since the operators are additionally suppressed by small couplings.

Among all the operators present in the effective Lagrangian at dimension six
\begin{equation}
\mathcal L^{(6)} =  \sum_i \frac{c_i}{\Lambda^2} Q_i
\,,
\end{equation}
that have been classified in \cite{Buchmuller:1985jz,Grzadkowski:2010es},
the ones contributing to both $b\to s\nu\bar\nu$ and $b\to s\ell^+\ell^-$ transitions are 
(omitting flavour indices)\footnote{Throughout, we use the notation $l_L$ to refer to the lepton doublet and $\ell$ ($=e,\mu,\tau$) to refer
to the lepton flavour in the basis where the charged lepton mass matrix is diagonal.},
\begin{align}
 Q_{Hq}^{(1)} &= i (\bar q_L \gamma_\mu q_L) H^\dagger D^\mu H
\,,&
 Q_{ql}^{(1)} &= (\bar q_L \gamma_\mu q_L) (\bar l_L\gamma^\mu l_L)
\,,\nonumber\\
 Q_{Hq}^{(3)} &= i (\bar q_L \gamma_\mu\tau^a q_L) H^\dagger D^\mu\tau_a H
\,,&
 Q_{ql}^{(3)} &= (\bar q_L \gamma_\mu \tau^a q_L) (\bar l_L\gamma^\mu \tau_al_L)
\,,\nonumber\\
 Q_{Hd} &= i (\bar d_R \gamma_\mu d_R) H^\dagger D^\mu H
\,,&
 Q_{dl} &= (\bar d_R \gamma_\mu d_R) (\bar l_L\gamma^\mu l_L)
\label{eq:ops}
\end{align}
and the ones contributing to $b\to s\ell^+\ell^-$ but {\em not} to $b\to s\nu\bar\nu$ are
\begin{align}
 Q_{de} &= (\bar d_R \gamma_\mu d_R) (\bar e_R\gamma^\mu e_R)
\,,&
 Q_{qe} &= (\bar q_L \gamma_\mu q_L) (\bar e_R\gamma^\mu e_R)
\,.
\label{eq:ops2}
\end{align}
For simplicity, we have omitted dipole operators, that are only relevant in semi-leptonic $b\to s\ell^+\ell^-$ processes at low dilepton invariant mass and 
in radiative decays, as well as scalar operators, that are only relevant in the $B_s\to\mu^+\mu^-$ decay (see also \cite{Alonso:2014csa}).

At low energies, after EWSB, the Wilson coefficients of these operators can be mapped onto the basis of the usual $\Delta F=1$ operators,
\begin{equation}
\mathcal H_\text{eff}^{\Delta F=1} = -\frac{4 G_F}{\sqrt{2}} V_{tb}V_{ts}^* \frac{e^2}{16\pi^2}\sum_i C_i \mathcal O_i,
\end{equation}
where the sum includes the operators $\mathcal O_{L,R}$ contributing to $b\to s\nu\bar\nu$ transitions as well as the following operators relevant for 
$b\to s\ell^+\ell^-$ transitions,
\begin{align}
\mathcal O_9^{(\prime)}&=
(\bar{s}  \gamma_{\mu} P_{L(R)} b)(  \bar{\ell} \gamma^\mu \ell)
\,,&
\mathcal O_{10}^{(\prime)}&=
(\bar{s}  \gamma_{\mu} P_{L(R)} b)(  \bar{\ell} \gamma^\mu\gamma_5 \ell)
\,.
\end{align}
Restoring flavour indices, working in the basis where the down-type quark mass matrix is diagonal and defining 
\begin{equation}
\widetilde{c}_k = \frac{(c_k)_{23}}{\Lambda^2}  \frac{\pi}{\sqrt{2}G_F \alpha V_{tb} V_{ts}^*} \approx \frac{(c_k)_{23}}{V_{tb}V_{ts}^*} \left(\frac{5 \,\text{TeV}}{\Lambda}\right)^2,
\label{eq:ctilde}
\end{equation}
one  can write
\begin{align}
C_L &= C_L^\text{SM} +  \widetilde{c}_{ql}^{(1)}- \widetilde{c}_{ql}^{(3)} +  \widetilde{c}_Z
\,,&
C_R &= \widetilde{c}_{dl} + \widetilde{c}_Z'
\label{eq:WC1}
\,,\\
C_9 &=C_9^\text{SM} + {\widetilde{c}_{qe}}+\widetilde{c}_{ql}^{(1)}+\widetilde{c}_{ql}^{(3)} -\zeta \,\widetilde{c}_Z
\,,&
C_9' &={\widetilde{c}_{de}} +\widetilde{c}_{dl} -\zeta \,\widetilde{c}_Z'
\label{eq:WC2}
\,,\\
C_{10} &=C_{10}^\text{SM} +{\widetilde{c}_{qe}}-\widetilde{c}_{ql}^{(1)} - \widetilde{c}_{ql}^{(3)} + \widetilde{c}_Z
\,,&
C_{10}' &= {\widetilde{c}_{de}} -\widetilde{c}_{dl}  + \widetilde{c}_Z'
\,,
\label{eq:WC3}
\end{align}
where
\begin{align}
\widetilde{c}_Z&=\tfrac{1}{2}(\widetilde{c}_{Hq}^{(1)}+\widetilde{c}_{Hq}^{(3)})
\,,
&
\widetilde{c}_Z'&=\tfrac{1}{2}\widetilde{c}_{Hd}
\,,
\end{align}
and $\zeta =1-4s_w^2\approx0.08$ is the accidentally small vector coupling of the $Z$ to charged leptons.
The operators with left-handed quarks also contribute to up-type FCNCs, but the constraints are weak in all cases.  The operators $Q_{Hq}^{(3)}$ and $Q_{ql}^{(3)}$ also contribute to flavour-changing charged currents, potentially modifying the the extraction of CKM elements from tree-level decays. We have checked that the constraints from FCNCs are stronger, barring cancellations.

We observe that the number of operators in the SM-EFT is in general larger
than in the low-energy effective Hamiltonian, so on a completely model-independent basis, no general correlations can be derived.
But in certain classes of NP scenarios, only a particular subset of operators is relevant and in this case correlations characteristic for this NP scenario are obtained.

While explicit models will be considered in the next section we illustrate 
general correlations on two examples.
We consider first the general case of $Z^\prime$ models in which a single $Z^\prime$ 
gauge boson dominates the scene. In this case, only the coefficients 
$ \widetilde{c}_{ql}^{(1)}$, $\widetilde{c}_{qe}$, $\widetilde{c}_{de}$ and 
$\widetilde{c}_{dl}$ are non-vanishing and we find
\begin{equation}\label{Zprime}
C_L^\text{NP}=\frac{C_9^\text{NP}-C_{10}^\text{NP}}{2}, \qquad 
C_R=\frac{C_9^\prime-C_{10}^\prime}{2},
\end{equation}
where with superscript NP we indicate the shift in Wilson coefficients due 
to new physics.

If NP contributions to the processes considered are fully dominated 
by induced FCNC couplings of the SM $Z$ boson -- this is the case e.g.\ in the MSSM and in models with partial compositeness -- then only the couplings 
$\widetilde{c}_Z$ and $\widetilde{c}_Z^\prime$ are non-vanishing. We find 
then
\begin{equation}
C_L=C_{10}^\text{NP}, \qquad C_9^\text{NP}=-\zeta C_{10}^\text{NP}
\end{equation}
and 
\begin{equation}
C_R=C_{10}^\prime, \qquad C_9^\prime=-\zeta C_{10}^\prime\, .
\end{equation}
The important difference from the $Z^\prime$ models is the flip of the sign in front of $C_{10}^\text{NP}$ in $C_L$ and similarly for $C_{10}^\prime$ in $C_R$.

Finally if the presence of $Z^\prime$ induces FCNC couplings of $Z$ through 
$Z-Z^\prime$ mixing, the relations in (\ref{Zprime}) are modified 
as follows
\begin{equation}\label{ZprimeZ}
C_L^\text{NP}=\frac{C_9^\text{NP}-C_{10}^\text{NP}}{2}+(3+\zeta)\frac{\widetilde{c}_Z}{2}, \qquad 
C_R=\frac{C_9^\prime-C_{10}^\prime}{2}+(3+\zeta)\frac{\widetilde{c}_Z^\prime}{2}.
\end{equation}
We observe that now the relations between $C_{L,R}$ and the Wilson coefficients 
 relevant for $b\to s\ell\ell$ depend on the size of $Z-Z^\prime$ mixing 
that generated non-vanishing coefficients $\widetilde{c}_Z$ and $\widetilde{c}_Z^\prime$.  This mixing is clearly model dependent and the resulting 
correlations can vary from model to model. We will illustrate this case 
in the next section by using the 331 models studied recently in \cite{Buras:2014yna}.

\subsection{Model-independent numerical analysis}\label{sec:smeftnum}
\subsubsection{General considerations}

The $b\to s\ell^+\ell^-$ Wilson coefficients $C_{9,10}^{(\prime)}$ are constrained by various experimental measurements, all of which are in agreement with the SM within uncertainties to date. In this section, we use these measurements to derive numerical bounds on the Wilson coefficients.
We make use of a global numerical analysis of NP in $b\to s \mu^+\mu^-$ transitions \cite{Altmannshofer:2011gn,Altmannshofer:2012ir,Altmannshofer:2013foa,Altmannshofer:2014rta}, including in particular
\begin{itemize}
 \item branching ratios of $B^{+,0}\to K^{+,0}\mu\mu$, $B^{+,0}\to K^{*+,0}\mu^+\mu^-$, $B_s\to\phi\mu^+\mu^-$ and $B_s\to\mu^+\mu^-$;
 \item the $B^0\to K^{*0}\mu^+\mu^-$ angular observables $A_\text{FB}$, $F_L$, $S_{3,4,5}$, and $A_9$.
\end{itemize}
Similar global analyses have been performed in \cite{Descotes-Genon:2013wba,Beaujean:2013soa}.

From~(\ref{eq:WC1})--(\ref{eq:WC3}), it is clear that in complete generality, the size of NP effects in $b\to s\nu\bar\nu$ is not constrained by the $b\to s\ell^+\ell^-$ measurements. First, the decays with charged leptons are only sensitive to the combination $(\widetilde{c}_{ql}^{(1)}+ \widetilde{c}_{ql}^{(3)})$, while the decays with neutrinos in the final state probe $(\widetilde{c}_{ql}^{(1)}- \widetilde{c}_{ql}^{(3)})$. Second, even if the Wilson coefficient $\widetilde{c}_{ql}^{(3)}$ vanishes, cancellations between the operators with left- and right-handed charged leptons can lead to small deviations from the SM in $b\to s\ell^+\ell^-$ transitions even when large effects are present in $b\to s\nu\bar\nu$. However, in concrete NP models, often only a subset of the operators are generated and the cancellations might happen only in fine-tuned corners of the parameter space. Therefore, we find it instructive to first look at the constraints on individual Wilson coefficients.

\subsubsection{Constraints on individual Wilson coefficients}

To this end, we construct a $\chi^2$ function in terms of the SM-EFT Wilson coefficients, including all the abovementioned observables. Varying the real or imaginary part of the individual Wilson coefficients, we obtain the following $2\sigma$ ($\Delta\chi^2=4$) ranges,
\begin{align}
\text{Re}(\tilde c_{ql}^{(1)}+\tilde c_{ql}^{(3)}) &\in [-0.84,-0.12]
\,,&
\text{Im}(\tilde c_{ql}^{(1)}+\tilde c_{ql}^{(3)}) &\in [-0.91,+0.89]
\,,\nonumber\\
\text{Re}\,\tilde c_{dl} &\in [-0.19,+0.33]
\,,&
\text{Im}\,\tilde c_{dl} &\in [-0.92,+0.89]
\,,\nonumber\\
\text{Re}\,\tilde c_{Z} &\in [-0.02,+1.03]
\,,&
\text{Im}\,\tilde c_{Z} &\in [-1.3,+1.3]
\,,\nonumber\\
\text{Re}\,\tilde c_{Z}' &\in [-0.53,+0.28]
\,,&
\text{Im}\,\tilde c_{Z}' &\in [-1.1,+1.3]
\,.
\label{eq:boundcZ}
\end{align}

We observe good agreement with the SM point $\tilde c_i=0$, except for the combination $\text{Re}(\tilde c_{ql}^{(1)}+\tilde c_{ql}^{(3)})$. This is due to the tensions recently observed in $B\to K^*\mu^+\mu^-$ angular observables and branching ratios of exclusive $b\to s$ transitions (cf. \cite{Aaij:2013qta,Descotes-Genon:2013wba,Altmannshofer:2013foa,Beaujean:2013soa,Lyon:2014hpa,Altmannshofer:2014rta}). We will see that, if due to NP, this tensions have an important impact on $b\to s\nu\bar\nu$ transitions.

Although the constraints on the imaginary parts of the Wilson coefficients are weaker than on the real parts, it can be easily seen from (\ref{eq:epsetadef}) and (\ref{eq:epseta-R}) that the impact of the imaginary parts on $\mathcal R_K$ and $\mathcal R_{K^*}$ is very small, as they do not interfere with the SM Wilson coefficient.

\begin{figure}[ptb]
\centering
\includegraphics[height=0.88\textheight]{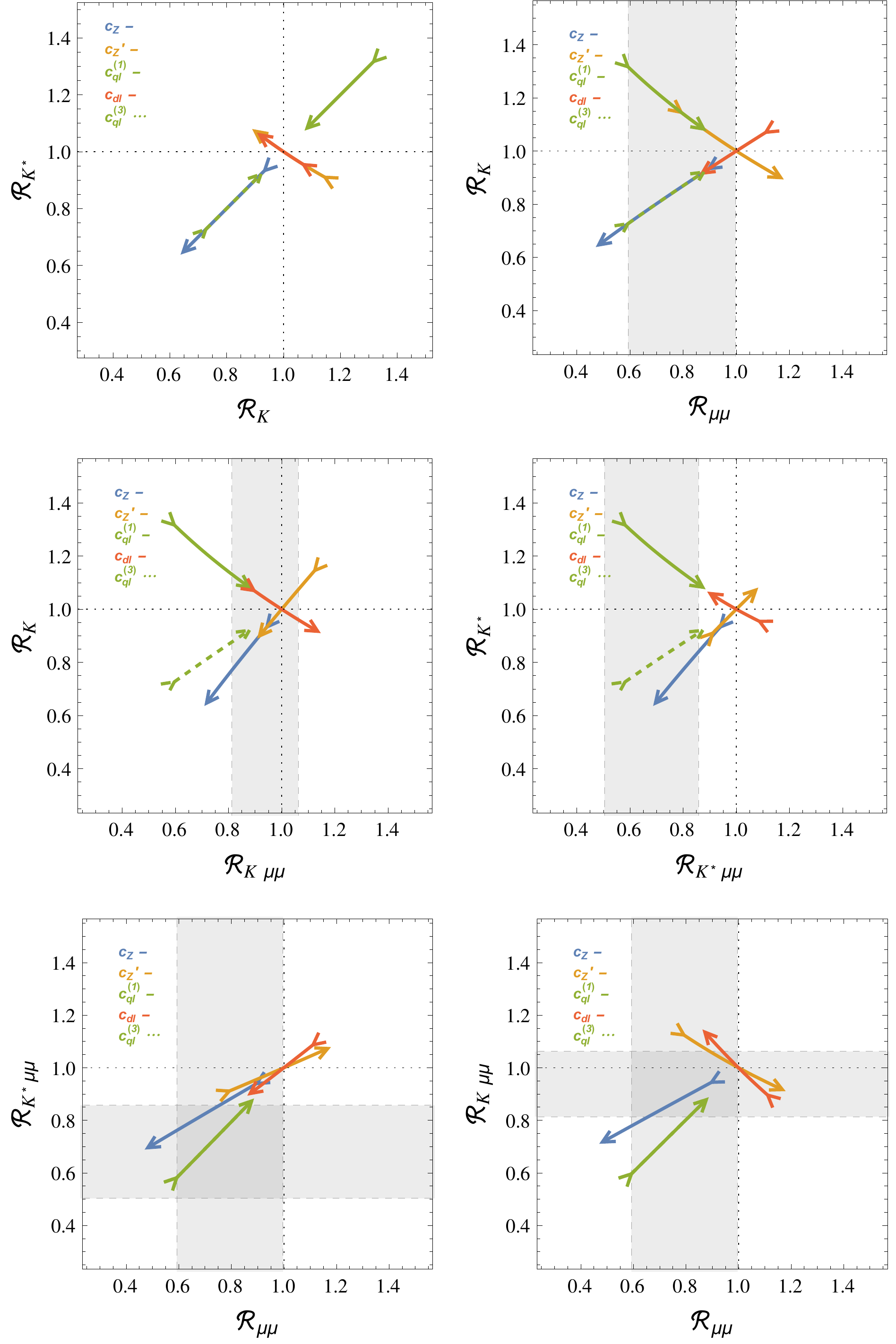}
\caption{Impact of the Wilson coefficients of the SM-EFT varied within their $2\sigma$ ranges allowed by the global fit to $b\to s\mu\mu$ data.
Blue: {\color{Pblue!80!black}\boldmath$\tilde c_Z$}, yellow: {\color{Pyellow!80!black}\boldmath$\tilde c_Z'$}, green: {\color{Pgreen!80!black}\boldmath$\tilde c_{ql}^{(1)}$} (solid) or
{\color{Pgreen!80!black}\boldmath$\tilde c_{ql}^{(3)}$} (dashed), red: {\color{Pred!80!black}\boldmath$\tilde c_{dl}$}. The arrows point from negative to positive (real) values of the Wilson coefficients. The shaded bands are the $1\sigma$ experimental measurements.}
\label{fig:Rplots}
\end{figure}

The impact of the real parts of the Wilson coefficients on the observables is visualized in figure~\ref{fig:Rplots}.
The colored arrows in these plots correspond to the $2\sigma$ allowed ranges of the individual Wilson coefficients as in (\ref{eq:boundcZ}), with the direction of the arrow pointing from negative to positive values for the $\tilde c_i$. The blue arrows correspond to $\tilde c_Z$, the yellow ones to $\tilde c_Z'$, the green ones to $\tilde c_{ql}^{(1)}$ (solid) or $\tilde c_{ql}^{(3)}$ (dashed), and the red ones to $\tilde c_{dl}$. Apart from $\mathcal R_K$ and $\mathcal R_{K^*}$, we also show the impact on the branching ratios of $B_s\to\mu^+\mu^-$ as well as of $B\to K\mu^+\mu^-$ and $B\to K^*\mu^+\mu^-$  at high $q^2$, defined as
\begin{align}
 \mathcal R_{\mu\mu} &= \frac{\text{BR}(B_s\to\mu^+\mu^-)}{\text{BR}(B_s\to\mu^+\mu^-)_\text{SM}}
 \,,\\
 \mathcal R_{K\mu\mu} &= \frac{\text{BR}(B^+\to K^+\mu^+\mu^-)^{[15,22]}}{\text{BR}(B^+\to K^+\mu^+\mu^-)^{[15,22]}_\text{SM}}
 \,,\\
 \mathcal R_{K^*\mu\mu} &= \frac{\text{BR}(B^0\to K^{*0}\mu^+\mu^-)^{[15,19]}}{\text{BR}(B^0\to K^{*0}\mu^+\mu^-)^{[15,19]}_\text{SM}}
 \,,
\end{align}
where the superscripts refer to the range in $q^2$ in GeV${}^2$. The shaded regions show the values allowed by direct experimental measurements at $1\sigma$. We make the following observations.
\begin{itemize}
\item The current data on $b\to s\mu^+\mu^-$ processes favour a negative $\tilde c_{ql}^{(1)}$, which implies an enhancement of $B\to K\nu\bar\nu$ and  $B\to K^*\nu\bar\nu$ by up to 30\% and a suppression of $B_s\to\mu^+\mu^-$, $B\to K\mu^+\mu^-$, and $B\to K^*\mu^+\mu^-$.
\item If there is NP in left-handed $Z$ penguins (i.e.\ in $\tilde c_{Z}$), current data imply a suppression of $B\to K\nu\bar\nu$ and  $B\to K^*\nu\bar\nu$ by up to 40\%.
\item With NP in right-handed currents, i.e.\ in $\tilde c_{dl}$ or $\tilde c_{Z}'$, $B\to K\nu\bar\nu$ and  $B\to K^*\nu\bar\nu$ can only be modified at the level of $\pm 10\%$.
\end{itemize}

\begin{figure}[tbp]
\centering
\includegraphics[width=0.5\textwidth]{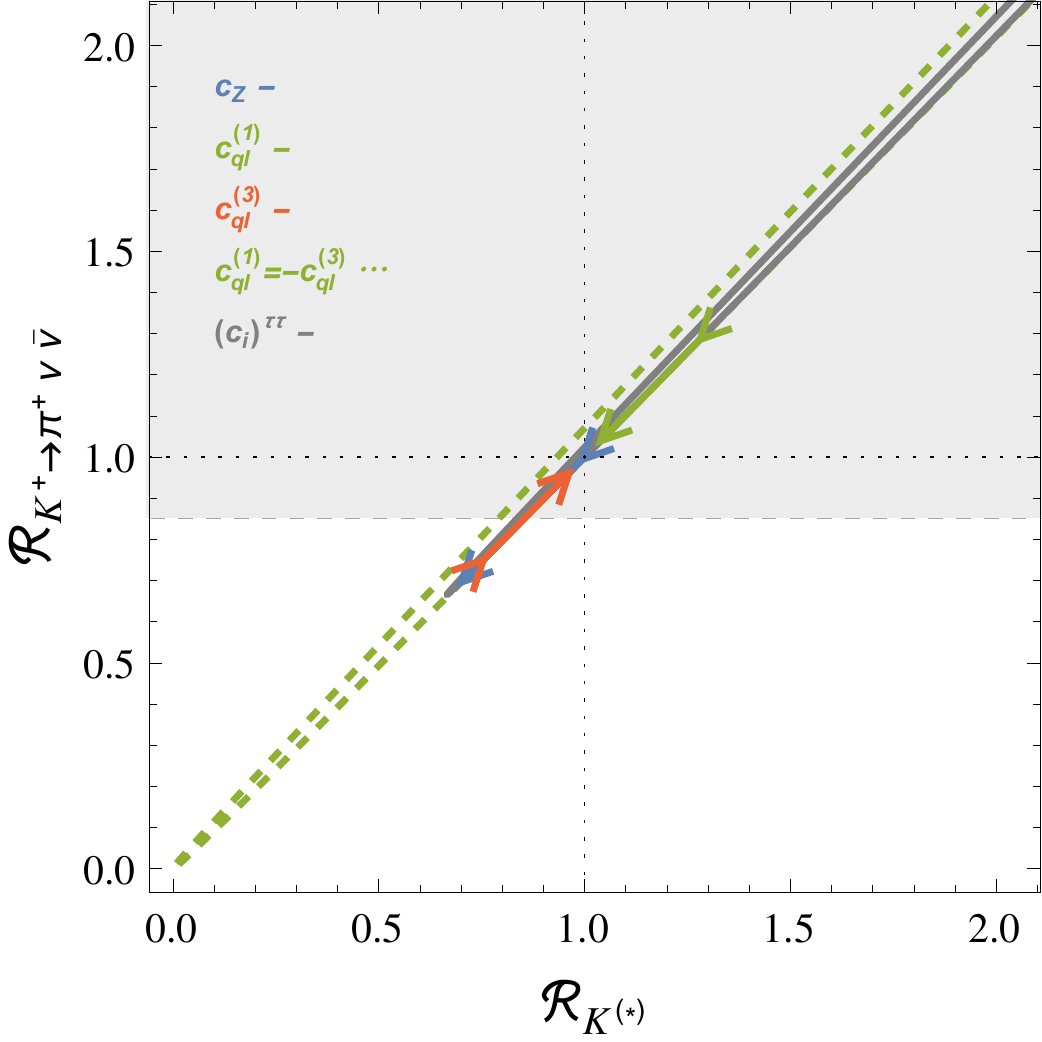}
\caption{Impact of the Wilson coefficients of the SM-EFT on $B\to K^{(*)}\nu\bar\nu$ vs. $K^+\to\pi^+\nu\bar\nu$, assuming MFV and LFU,  varied within their $2\sigma$ ranges allowed by the global fit to $b\to s\mu\mu$ data.
Blue: {\color{Pblue!80!black}\boldmath$\tilde c_Z$}, green: {\color{Pgreen!80!black}\boldmath$\tilde c_{ql}^{(1)}$}, red: {\color{Pred!80!black}\boldmath$\tilde c_{ql}^{(3)}$}. The dashed green line shows the case $\tilde c_{ql}^{(1)}=-\tilde c_{ql}^{(3)}$, where $b\to s\mu\mu$ constraints are ineffective.
The gray line corresponds to NP in operators with tau neutrinos only, where $\mathcal R_K^{(*)}>2/3$ (see section~\ref{sec:lfu}).
The shaded band is the $1\sigma$ experimental measurement \cite{Artamonov:2008qb}.}
\label{fig:Rplot-MFV}
\end{figure}

\subsubsection{The case of Minimal Flavour Violation (MFV)}
An interesting special case is MFV, where only the Wilson coefficients $\tilde c_{ql}^{(1,3)}$, $\tilde c_{qe}$, and $\tilde c_{Z}$ are 
allowed and are real. Moreover, the same Wilson coefficients $\tilde c_i$ also enter $s\to d\nu\bar\nu$ transitions implying strict correlations between the latter 
processes and the ones considered here \cite{Buras:2001af}. As the precise rate for $K^+\to\pi^+\nu\bar\nu$ is expected to be known before the ones for 
$B\to K^{(*)}\nu\bar\nu$, additional constraints on the latter decays will follow in addition to the ones from $b\to s\ell^+\ell^-$ processes considered by us.
In figure~\ref{fig:Rplot-MFV}, we show the correlation between $B\to K\nu\bar\nu$ and $K^+\to\pi^+\nu\bar\nu$, normalized to their SM values. As  
$\mathcal R_{K^*}= \mathcal R_{K}$ 
in this framework, the same correlation applies to $B\to K^*\nu\bar\nu$.

When varied separately, with $b\to s\ell^+\ell^-$ constraints taken into account, the above Wilson coefficients lead to deviations of up to $\pm30\%$ from the SM branching ratios. 
 Very large effects can be obtained in principle in models where $\tilde c_{ql}^{(1)}=-\tilde c_{ql}^{(3)}$, so contributions to $b\to s\ell^+\ell^-$ 
processes vanish and only present weak constraint from $K^+\to\pi^+\nu\bar\nu$  play a role. Once the rate for $K^+\to\pi^+\nu\bar\nu$ will be experimentally known with
a high precision, it will be possible to obtain the allowed region 
for $\mathcal R_{K^*}= \mathcal R_{K}$ with the same precision under the assumption of MFV, thereby selecting the favourite MFV models. Eventually one would hope to find the experimental allowed 
region in the plot in figure~\ref{fig:Rplot-MFV} to be outside the straight line. This would be an important signal of non-MFV interactions at work, in particular in 
view of the fact that all processes involved belong to the theoretically cleanest in the field of rare decays. Additional important information will come from 
CP-violating decay $K_L\to\pi^0\nu\bar\nu$ but this will take more time.

\subsubsection{{\em Z} contributions vs. 4-fermion operators}

\begin{figure}[tb]
\centering
\includegraphics[width=0.65\textwidth]{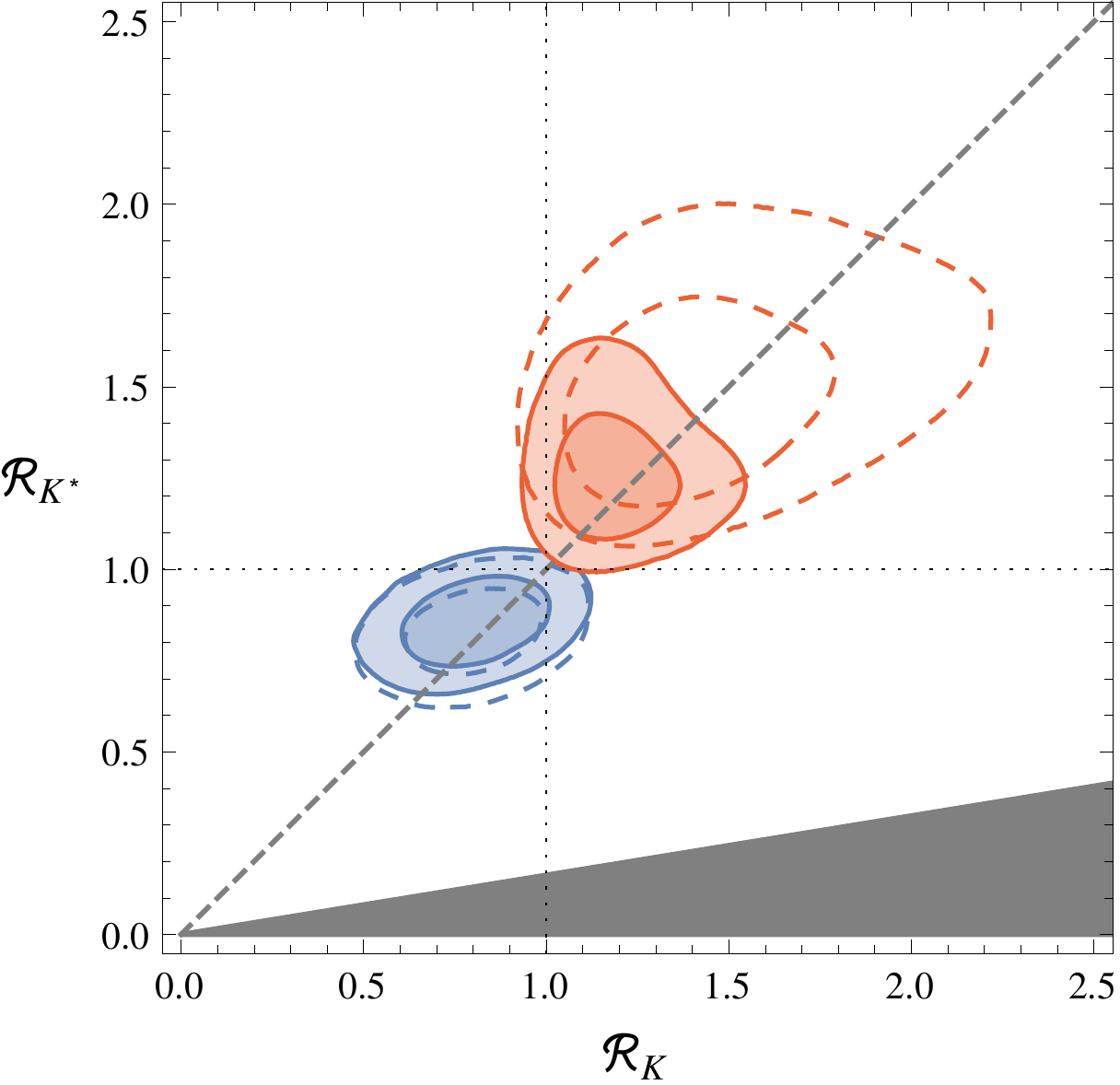}%
\caption{Constraints on the branching ratios of $B\to K\nu\bar\nu$ and $B\to K^*\nu\bar\nu$ normalized to their SM values from a global analysis of $b\to s\mu^+\mu^-$
processes.
Blue: assuming NP in in flavour-changing $Z$ couplings only, with real (solid) or complex (dashed) Wilson coefficients.
Red: assuming NP in 4-fermion operators only (as happens if the NP contributions are dominated by a SM-singlet $Z'$, see sec.~\ref{sec:genzp}.), with real (solid) or complex (dashed) Wilson coefficients, assuming LFU.}
\label{fig:BR-Z4f}
\end{figure}

Even if NP contributes to more than one coefficient, there are scenarios in which the NP effects in $b\to s\nu\bar\nu$ are quite limited. We now consider the case where NP contributes only through (tree-level or loop-induced) flavour-changing $Z$ couplings, i.e. through 
$\tilde c_{Z}$ and $\tilde c_{Z}'$. To determine the allowed size of the observables, we performed a Markov Chain Monte Carlo scan of the Wilson coefficients $\tilde c_{Z}$ and $\tilde c_{Z}'$, assuming flat priors for them.
The resulting allowed region in the $\mathcal R_K$-$\mathcal R_{K^*}$ plane is shown in figure~\ref{fig:BR-Z4f} for real Wilson coefficients (solid blue contours, corresponding to contributions aligned in phase with the SM) as well as for complex Wilson coefficients (dashed blue contours, corresponding to new sources of CP violation).

An orthogonal possibility is to have NP contributions only in Wilson coefficients of 4-fermion operators\footnote{Using this terminology we distinguish these contributions from 
the ones coming from $Z$ contributions even if the latter ones generate at low energies 4-fermion operators.}, 
i.e. in $\tilde c_{ql}^{(1)}$, $\tilde c_{qe}$, $\tilde c_{dl}$, and $\tilde c_{de}$ (but not $\tilde c_{ql}^{(3)}$, which is unconstrained as discussed above),
which is what happens if the new physics contributions are dominated by a single SM-singlet $Z'$ gauge boson, as will be discussed in section~\ref{sec:genzp}.
Again, we show the allowed region in the $\mathcal R_K$-$\mathcal R_{K^*}$ plane for real or complex Wilson coefficients as solid and dashed red contours in figure~\ref{fig:BR-Z4f}.

The striking feature of figure~\ref{fig:BR-Z4f} is that the current data on $b\to s\mu^+\mu^-$ transitions show some tension with the SM predictions; solving these tensions via $Z$ penguins or via four-fermion operators in the SM-EFT leads to very different predictions for $B\to K\nu\bar\nu$ and $B\to K^*\nu\bar\nu$. In the former case, a suppression is predicted, in the latter an enhancement (that, for complex coefficients, even comes close to saturating the current experimental bound of $\mathcal R_K < 4.3$).
Needless to say, the tensions in $b\to s\mu^+\mu^-$ data might be due to statistical fluctuations or underestimated uncertainties, but we find it 
interesting that a measurement of $B\to K\nu\bar\nu$ and $B\to K^*\nu\bar\nu$ could help pin down the source of NP contributions.

\subsection{Beyond lepton flavour universality}\label{sec:lfu}

So far, we have assumed the Wilson coefficients to be independent of the lepton flavour. In general however, they could be different for different lepton flavours while still being lepton flavour conserving -- we call this lepton flavour non\hyp{}universality (LFNU) -- and there could even be lepton flavour violation (LFV). In this section, we discuss the implications of these two scenarios for $b\to s\nu\bar\nu$.

\subsubsection{LFNU}

In the case of LFNU, the Wilson coefficients in the low-energy effective theory get an index $\ell=e,\mu,\tau$ distinguishing the neutrino flavour,
\begin{align}
\mathcal{O}_{L}^\ell &=\frac{e^2}{16\pi^2}
(\bar{s}  \gamma_{\mu} P_L b)(  \bar{\nu_\ell} \gamma^\mu(1- \gamma_5) \nu_\ell)
\,,&
\mathcal{O}_{R}^\ell &=\frac{e^2}{16\pi^2}(\bar{s}  \gamma_{\mu} P_R b)(  \bar{\nu_\ell} \gamma^\mu(1- \gamma_5) \nu_\ell)
\,.
\end{align}
Defining
\begin{equation}  \label{eq:epsetadef2}
 \epsilon_\ell = \frac{\sqrt{ |C_L^\ell|^2 + |C_R^\ell|^2}}{|C_L^\text{SM}|}~, \qquad
 \eta_\ell = \frac{-\text{Re}\left(C_L^\ell C_R^{\ell *}\right)}{|C_L^\ell|^2 + |C_R^\ell|^2}~,
\end{equation}
in analogy to (\ref{eq:epsetadef2}), the generalization of (\ref{eq:epseta-R}) reads

\begin{align}
 \mathcal{R}_K \equiv \frac{\mathcal{B}_K}{\mathcal{B}_K^\text{SM}}   & = 
  \frac{1}{3}\sum_\ell (1 - 2\,\eta_\ell)\epsilon_\ell^2
 \,, \\
 \mathcal{R}_{K^*} \equiv \frac{\mathcal{B}_{K^*}}{\mathcal{B}_{K^*}^\text{SM}} 
  & =
  \frac{1}{3}\sum_\ell (1 +  \kappa_\eta \eta_\ell)\epsilon_\ell^2
  \,, \\
\label{eq:epseta-LFNU}
 \mathcal{R}_{F_L} \equiv \frac{F_L}{F_L^\text{SM}} 
 & =  
 \frac{\sum_\ell \epsilon_\ell^2(1 + 2 \,\eta_\ell)}{\sum_\ell \epsilon_\ell^2(1 + \kappa_\eta \eta_\ell)}
 \,.
\end{align}
One can check that the model-independent relations (\ref{eq:FLtest}) and (\ref{eq:Xstest}) still hold.
If NP only contributes to operators involving one of the three lepton flavours, one obtains model-independent lower bounds $\mathcal R_K>2/3$, $\mathcal R_{K^*}>2/3$.

Also the SM-EFT analysis of section~\ref{sec:smeft} can be generalized to the case of LFNU. Since the effective Wilson coefficients $\tilde c_Z^{(\prime)}$ arise
from operators not involving lepton fields at all, their effects are always lepton flavour universal and the bounds in (\ref{eq:boundcZ}) still apply.

The Wilson coefficient of the four-fermion operators instead become lepton flavour dependent and we discuss effects due to operators only involving electrons, 
muons or taus (and their respective neutrinos) in turn.\footnote{As in the LFU case, NP entering the combination $(\widetilde{c}_{ql}^{(1)})^\ell- (\widetilde{c}_{ql}^{(3)})^\ell$ but not $(\widetilde{c}_{ql}^{(1)})^\ell+ (\widetilde{c}_{ql}^{(3)})^\ell$ is unconstrained by $b\to s\ell^+\ell^-$ processes and can in principle give rise to large effects in $b\to s\nu\bar\nu$.}

\paragraph{\boldmath$\ell=\mu$}
The constraints on the Wilson coefficients of the four-fermion operators involving muon fields, $(\widetilde{c}_{ql}^{(1)})^\mu+ (\widetilde{c}_{ql}^{(3)})^\mu$, $(c_{qe})^\mu$, $(c_{dl})^\mu$, and  $(c_{de})^\mu$, are the same as the LFU scenario considered in section~\ref{sec:smeftnum}, since we considered only constraints from decays involving muons in the final state. However, the deviations from the SM in $B\to K^{(*)}\nu\bar\nu$ are now a factor of $\frac{1}{3}$ smaller, since only muon neutrinos contribute. 
The allowed region in the $\mathcal R_K$-$\mathcal R_{K^*}$ plane in figure~\ref{fig:BR-Z4f}, shown in red, would thus shrink by this factor (in a geometrically similar way, i.e.\ without changing its shape).
We note that the case with NP in $\ell=\mu$ only is particularly interesting in view 
of the recent measurement by LHCb of the ratio of the $B\to K e^+e^-$ to $B\to K \mu^+\mu^-$ branching ratios at low $q^2$, found to deviate from LFU by $2.6\sigma$
\cite{Aaij:2014ora} (see also \cite{Hiller:2014yaa}).

\paragraph{\boldmath$\ell=\tau$}
For four-fermion operators involving taus, constraints from semi-leptonic FCNCs are more than two orders of magnitude weaker than for muons \cite{Bobeth:2011st}. Consequently, the effects in $b\to s\nu\bar\nu$ decays could easily saturate the experimental upper bounds.
In fact, the current upper bounds on the $B\to K^{(*)} \nu\bar\nu$ branching ratios represent the most stringent bounds on FCNC operators in the SM-EFT involving left-handed tau leptons and limit the size of NP effects that can be generated in $b\to s\tau^+\tau^-$ decays from these operators.

\begin{figure}[tbp]
\centering
\includegraphics[width=\textwidth]{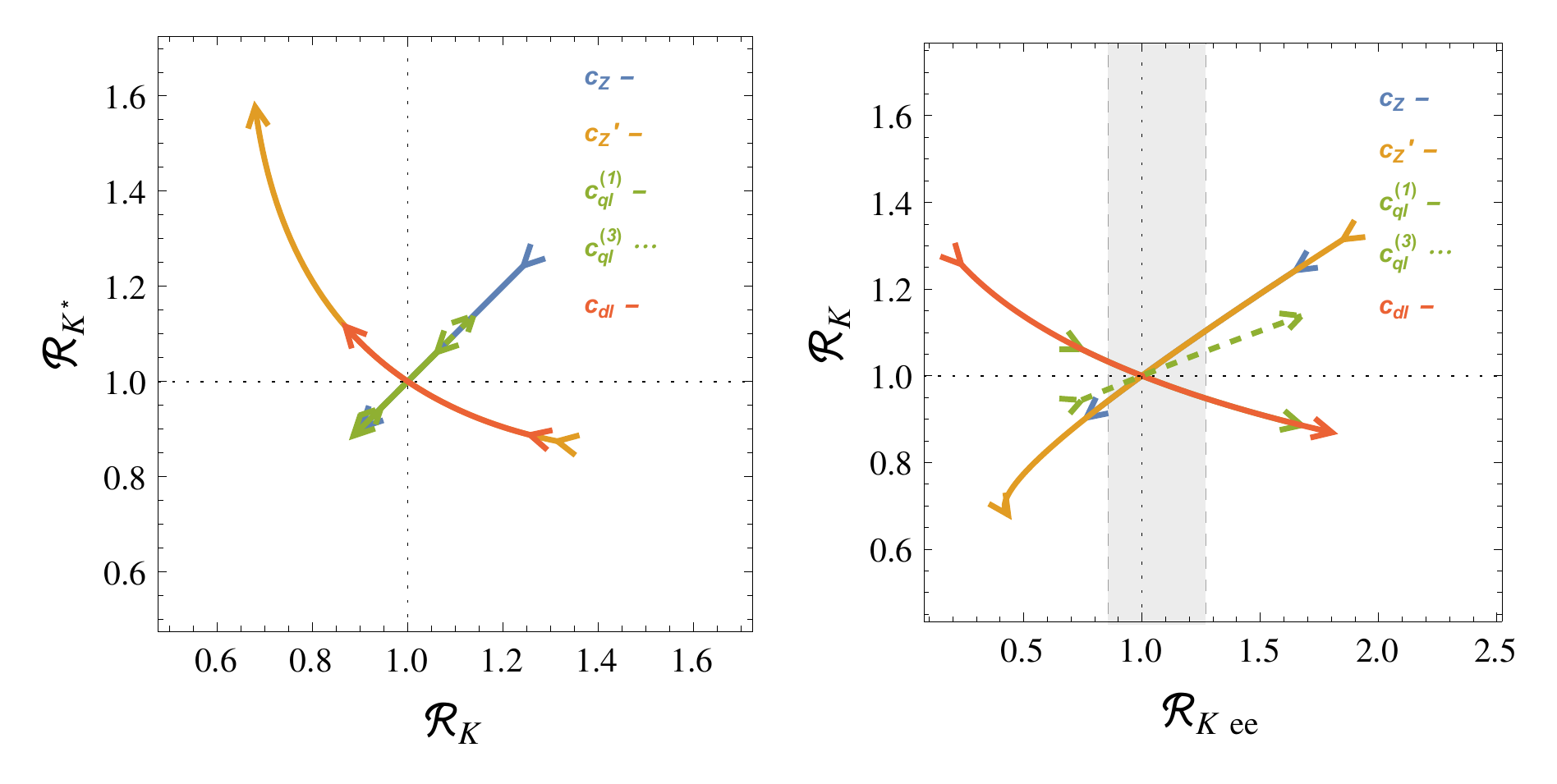}
\caption{Impact of the Wilson coefficients of the SM-EFT varied within their $2\sigma$ ranges,
assuming coefficients with electrons only, as allowed by a fit to $b\to se^+e^-$ data.
Colors as in figure~\ref{fig:Rplots}.}
\label{fig:Rplots-ee}
\end{figure}

\paragraph{\boldmath$\ell=e$}
In the case of four-fermion operators involving electron fields, the only constraints at present are 
the branching ratio of the inclusive $B\to X_se^+e^-$ decay measured by BaBar \cite{Lees:2013nxa} as well the branching ratio of $B^+\to K^+ e^+e^-$ measured recently by LHCb \cite{Aaij:2014ora}. Using these measurements, we obtain the following $2\sigma$ allowed ranges for the Wilson coefficients,
\begin{align}
\text{Re}(\tilde c_{ql}^{(1)}+\tilde c_{ql}^{(3)}) &\in [-0.42,+0.92]
&
\text{Im}(\tilde c_{ql}^{(1)}+\tilde c_{ql}^{(3)}) &\in [-2.9,+2.9]
\\
\text{Re}\,\tilde c_{dl} &\in [-0.83,+0.91]
&
\text{Im}\,\tilde c_{dl} &\in [-2.9,+2.9]
\\
\text{Re}\,\tilde c_{Z} &\in [-1.6,+0.8]
&
\text{Im}\,\tilde c_{Z} &\in [-4.0,+4.0]
\\
\text{Re}\,\tilde c_{Z}' &\in [-1.6,+1.8]
&
\text{Im}\,\tilde c_{Z}' &\in [-4.0,+4.0]
\end{align}
These bounds are significantly looser than in the LFU case (\ref{eq:boundcZ}), so in spite of the fact that the effects in the $B\to K^{(*)}\nu\bar\nu$ branching ratios are a factor of 3 smaller, since only one neutrino flavour contributes, the allowed effects can be bigger. This is visualized in figure \ref{fig:Rplots-ee}, showing in analogy to figure~\ref{fig:Rplots} the correlation between $\mathcal R_K$ and $\mathcal R_{K^*}$ as well as between $\mathcal R_K$ and the low-$q^2$ branching ratio
\begin{equation}
 \mathcal R_{Kee} = \frac{\text{BR}(B^+\to K^+e^+e^-)^{[1,6]}}{\text{BR}(B^+\to K^+e^+e^-)^{[1,6]}_\text{SM}}
 \,.
\end{equation}
 
\subsubsection{LFV}

In general, the operators in the low-energy Hamiltonian could violate lepton flavour,
\begin{align}
\mathcal{O}_{L}^{ij} &=\frac{e^2}{16\pi^2}
(\bar{s}  \gamma_{\mu} P_L b)(  \bar{\nu_i} \gamma^\mu(1- \gamma_5) \nu_j)
\,,&
\mathcal{O}_{R}^{ij} &=\frac{e^2}{16\pi^2}(\bar{s}  \gamma_{\mu} P_R b)(  \bar{\nu_i} \gamma^\mu(1- \gamma_5) \nu_j)
\,,
\end{align}
with $i\neq j$. 
In analogy to the case of LFNU, one can now define
\begin{equation}  \label{eq:epsetadef3}
 \epsilon_{ij} = \frac{\sqrt{ |C_L^{ij}|^2 + |C_R^{ij}|^2}}{|C_L^\text{SM}|}~, \qquad
 \eta_{ij} = \frac{-\text{Re}\left(C_L^{ij} C_R^{{ij} *}\right)}{|C_L^{ij}|^2 + |C_R^{ij}|^2}~,
\end{equation}
and (\ref{eq:epseta-LFNU}) still hold with the replacements $\epsilon_\ell,\eta_\ell$ $\to$ $\epsilon_{ij},\eta_{ij}$. Note that 
in the SM, $\epsilon_{ij}=0$ for $i\neq j$.
It is now easy to see that the model-independent relations (\ref{eq:FLtest}) and (\ref{eq:Xstest}) still hold.

Only few searches for LFV $B$ decays exist. The most stringent bound on LFV operators in the SM-EFT comes from the recent search for $B_s\to e\mu$  by the LHCb collaboration \cite{Aaij:2013cby}, finding
\begin{equation}
 \text{BR}(B_s\to e^\pm\mu^\mp) < 1.4\times 10^{-8}
\end{equation}
at 95\% confidence level. This implies
\begin{equation}
|(C_9)^{\mu e}-(C_9')^{\mu e}|^2+|(C_{10})^{\mu e}-(C_{10}')^{\mu e}|^2 < 16.6^2
\label{eq:Bsemu1}
\end{equation}
and likewise for the $(C_i^{(\prime)})^{e \mu}$. These Wilson coefficients are the obvious generalizations of (\ref{eq:WC2}) and (\ref{eq:WC3}) for the LFV case and
they vanish in the SM.
We are not aware of any existing bounds on LFV operators involving tau leptons.

We can now consider as an example a NP effect only in $(\widetilde{c}_{ql}^{(1)})^{e\mu}$.  As there is no interference between the SM and NP contribution in 
this case, we obtain
\begin{equation}
\mathcal R_K=\mathcal R_{K^*}=1+\frac{1}{3}\frac{|(\widetilde{c}_{ql}^{(1)})^{e\mu}|^2}{|C_L^\text{SM}|^2}\,.
\end{equation}
Using next
\begin{equation}
 C_L^\text{SM}=-6.35, \qquad |(\widetilde{c}_{ql}^{(1)})^{e\mu}|\le 11.7 \,,
\end{equation}
with the latter bound following from (\ref{eq:Bsemu1}), we find 
\begin{equation}
1\le \mathcal R_K=\mathcal R_{K^*} \le 2.14\, .
\end{equation}

To summarize, a NP contribution to $B\to K^{(*)}\nu\bar\nu$ that is purely LFV
\begin{itemize}
 \item always leads to an enhancement of the branching ratios,
 \item can lead to a factor of 2 enhancement of $\mathcal R_{K^{(*)}}$ for $ij=e\mu$, constrained by the search for $B_s\to e^\pm\mu^\mp$,
 \item can saturate the experimental bounds if tau neutrinos are involved.
\end{itemize}

\section{Specific new physics models}\label{sec:4}

\subsection[General \texorpdfstring{$Z'$}{Z'} Models]{General $\bldm{Z'}$ Models}\label{sec:genzp}

We will next consider general $Z^\prime$ models assuming that NP contributions 
are dominated by the tree-level exchange of a heavy neutral gauge boson with 
mass $M_{Z^\prime}$ that transforms as a singlet under $SU(2)_L$. The recent detailed analyses of FCNCs in these models 
can be found in \cite{Buras:2012jb,Buras:2013qja,Buras:2014zga}. We will follow the 
notation of $Z^\prime$ couplings in these papers,
\begin{equation}
\mathcal L \supset \bar f_i \gamma^\mu \left[ \Delta^{f_if_j}_L(Z') P_L + \Delta^{f_if_j}_R(Z') P_R \right] f_j \,Z'_\mu
\,,
\end{equation}
and recall
that $SU(2)_L$ symmetry implies
$\Delta_{L}^{\nu\bar\nu}(Z^\prime) =\Delta_{L}^{\ell\ell}(Z^\prime)$. The quark couplings are in general complex whereas the leptonic ones are 
assumed to be real. 
This results in the following tree-level contributions to the Wilson coefficients in the SM-EFT,
\begin{align}
\tilde c_{ql}^{(1)} &= -\frac{\Delta^{sb}_L\Delta^{\ell\ell}_L}{V_{tb}V_{ts}^*}\left[\frac{5\,\text{TeV}}{M_{Z'}}\right]^2
\,,&
\tilde c_{dl} &= -\frac{\Delta^{sb}_R\Delta^{\ell\ell}_L}{V_{tb}V_{ts}^*}\left[\frac{5\,\text{TeV}}{M_{Z'}}\right]^2
\,,\\
\tilde c_{qe} &= -\frac{\Delta^{sb}_L\Delta^{\ell\ell}_R}{V_{tb}V_{ts}^*}\left[\frac{5\,\text{TeV}}{M_{Z'}}\right]^2
\,,&
\tilde c_{de} &= -\frac{\Delta^{sb}_R\Delta^{\ell\ell}_R}{V_{tb}V_{ts}^*}\left[\frac{5\,\text{TeV}}{M_{Z'}}\right]^2
\,,
\end{align}
where the 5\,TeV stem from~(\ref{eq:ctilde}).
Here we have again assumed LFU. In that case, the lepton couplings are constrained from LEP2 searches for contact interactions \cite{Acciarri:2000uh,Abbiendi:2003dh},
\begin{align}
\frac{|\Delta^{\ell\ell}_L|}{M_{Z'}} &< \frac{0.41}{\text{TeV}}
\,,
&
\frac{|\Delta^{\ell\ell}_R|}{M_{Z'}} &< \frac{0.44}{\text{TeV}}
\,.
\end{align}
The quark couplings are in general complex. The mass difference in $B_s$-$\bar B_s$ mixing leads to the constraint
\begin{align}
\frac{1}{M_{Z'}^2}\left|(\Delta^{bs}_L)^2+(\Delta^{bs}_R)^2-8.6\,\Delta^{bs}_L\Delta^{bs}_R \right|
< \left(\frac{0.004}{\text{TeV}}\right)^2
\,,
\end{align}
where the numerical factor $8.6$ corresponds to $M_{Z^\prime}=5\,\text{TeV}$. It increases logarithmically with $M_{Z^\prime}$ reaching $10.0$ for  $M_{Z^\prime}=20\,\text{TeV}$. 
Details can be found in \cite{Buras:2014zga}.
Similarly, the measurement of the $B_s$ mixing phase constrains the argument of the combination of couplings.
Since we do not specify the flavour-conserving couplings to first and second generation quarks, we do not need to consider direct LHC bounds or bounds from atomic parity violation.

 Allowing the flavour-violating couplings to quarks to be complex, in principle one can arrange for cancellations in the $B_s$ mixing constraints and
 obtain large effects in $B\to K^{(*)}\nu\bar\nu$. Barring such fine-tuned scenarios, it is more instructive to consider several cases for the ratio 
 between left- and right-handed flavour-changing couplings. We will consider four cases:
the scenario in 
which only LH quark couplings are present (LHS), the one with only RH couplings 
(RHS), the one with LH and RH couplings being equal (LRS) and one with 
these couplings differing by sign (ALR). We will use the following colour 
coding for them
\be\label{colourcoding}
 \text{LHS}=\text{(red)},\quad  \text{RHS}=\text{(blue)},\quad  \text{LRS}=\text{(green)},\quad  \text{ALRS}=\text{(yellow)}.
\ee
Since we are mainly interested in $b\to s\nu\bar\nu$ transitions, we consider only $Z'$ couplings to left-handed leptons,
\be\label{leptonhigh}
\Delta_R^{\nu\bar\nu}(Z^\prime) =\Delta_R^{\ell\ell}(Z^\prime)=0\,.
\ee

The relevant formulae for processes of interest in terms of these couplings 
are collected in \cite{Buras:2012jb,Buras:2013qja} and we will not repeat them 
here.
The $\Delta F=2$ constraint has been incorporated through the conditions 
\be\label{DF2c}
-0.14 \le S_{\psi\phi} \le 0.14, \qquad 0.9\le C_{B_s}\equiv
\frac{\Delta M_s}{\Delta M_s^{\text{SM}}}\le 1.1 \,.
\ee

\begin{figure}[ptb]
\centering
\includegraphics[height=0.9\textheight]{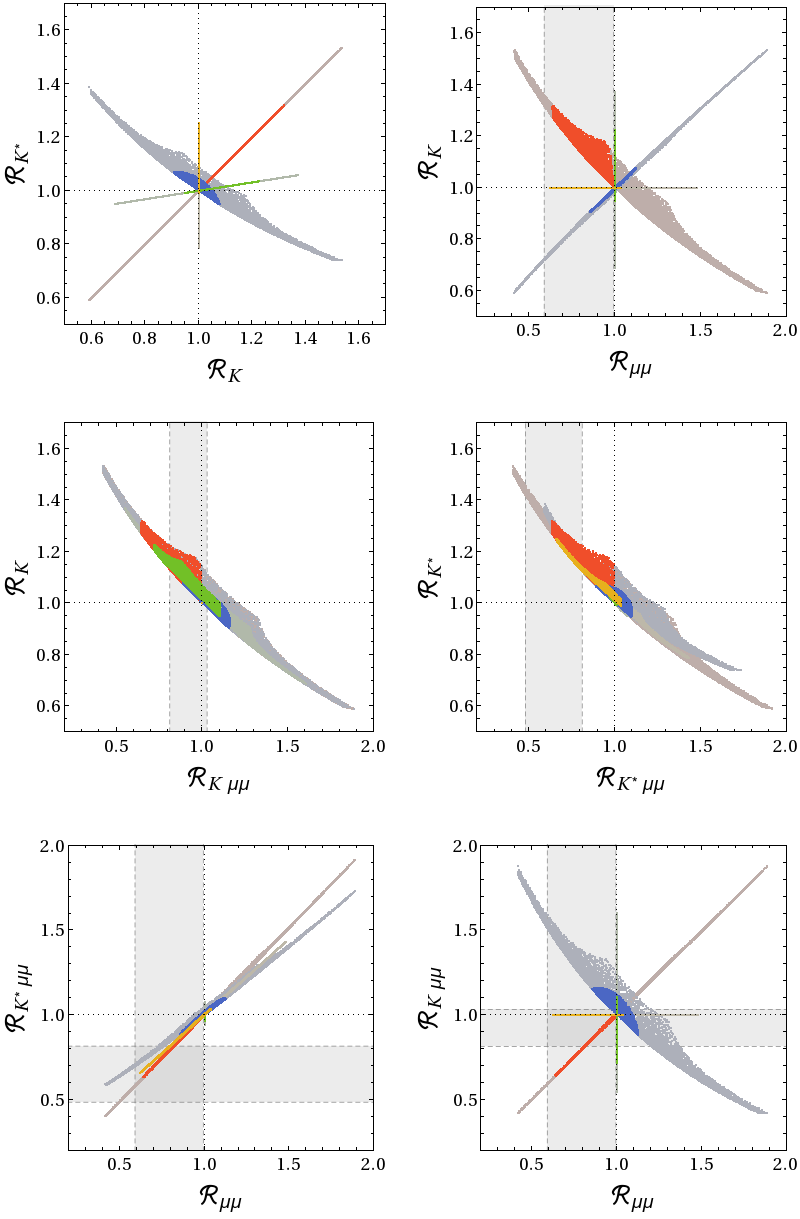}
\caption{Various correlations between observables in  LHS (red), RHS (blue), LRS (green), ALRS (yellow), assuming LFU and $\Delta_R^{\nu\nu} = \Delta_R^{\ell\ell}=0$. All points satisfy  $0.9\leq C_{B_s}\leq 1.1$, $-0.14\leq S_{\psi\phi}\leq 0.14$.
Grey  regions are disfavoured at $2\sigma$ by $b\to s\mu^+\mu^-$ constraints.}
\label{fig:Zprime}
\end{figure}

In figure~\ref{fig:Zprime} we show the correlations between the same observables
as in figure~\ref{fig:Rplots}.
All points fulfill the $\Delta F=2$ constraints. Some of the points passing $\Delta F=2$ constraints are disfavoured by the global analysis 
of $b\to s\mu^+\mu^-$ data. These are shown as grey regions in the figure.
These plots show in a spectacular manner how different 
scenarios for couplings can be distinguished through correlations. 
Qualitative understanding of these plots can be gained by inspecting 
DNA charts in  \cite{Buras:2013ooa} paying attention to the signs 
of the leptonic couplings as explained there.

Here we just want to emphasize most important points. The choice of the couplings in (\ref{leptonhigh}) corresponds to a very simple structure of the Wilson coefficients in 
the effective theory approach
\begin{align}
C_L &= C_L^\text{SM} +  \widetilde{c}_{ql}^{(1)}
\,,&
C_R &= \widetilde{c}_{dl} 
\label{eq:WC1SU(2)}
\,,\\
C_9 &=C_9^\text{SM} + \widetilde{c}_{ql}^{(1)}
\,,&
C_9' &=\widetilde{c}_{dl} 
\,,\\
C_{10} &=C_{10}^\text{SM} -\widetilde{c}_{ql}^{(1)} 
\,,&
C_{10}' &=  -\widetilde{c}_{dl}  
\,,
\label{eq:WC3SU(2)}
\end{align}
implying correlations between Wilson coefficients
\begin{equation}
 C_L^\text{NP}=C_9^\text{NP}=-C_{10}^\text{NP}, \qquad C_R=C_9^\prime=-C_{10}^\prime.
\end{equation}
The relation $C_9^\text{NP}=-C_{10}^\text{NP} <0$ has been recently advocated in \cite{Hiller:2014yaa} as a simple description of the present data on 
$b\to s \mu^+\mu^-$ transitions.

As a result of these correlations we have the following unique predictions seen in figure~\ref{fig:Zprime}:
\begin{itemize}
\item
In the LHS $\mathcal{R}_K$ and $\mathcal{R}_K^{*}$ are correlated with each 
other, whereas in the RHS they are anti-correlated.
\item
The suppression of the rate for $B_s\to\mu^+\mu^-$ ($\mathcal{R}_{\mu\mu}< 1$) relative to its SM value implies the enhancement of $\mathcal{R}_K$ in the LHS and 
its suppression in the RHS.
\item
 $\mathcal{R}_K$ is anti-correlated with  $\mathcal{R}_{K\mu\mu}$ independently  of scenario considered. The same applies to the relation between 
  $\mathcal{R}_K^{*}$  and  $\mathcal{R}_{K^{*}\mu\mu}$. 
\item
In accordance with the present data, $\mathcal{R}_{\mu\mu}< 1$ implies 
 $\mathcal{R}_{K^{*}\mu\mu}<1$ in all scenarios considered.
\item 
On the other hand the simultaneous suppression of  $\mathcal{R}_{\mu\mu}< 1$ 
and  $\mathcal{R}_{K\mu\mu}<1$ observed in the data favours the LHS and 
strongly disfavours the RHS.
\end{itemize}

So far, we have assumed LFU. Here, we briefly
comment on the $Z'$ model with gauged muon minus tau lepton number $(L_\mu-L_\tau)$ proposed recently in \cite{Altmannshofer:2014cfa} to address a number of 
tensions observed in $b\to s\mu^+\mu^-$ transitions. In this model, one has
\begin{align}
(\tilde c_{qe})^{\mu\mu}&=(\tilde c_{ql}^{(1)})^{\mu\mu} = -(\tilde c_{qe})^{\tau\tau}= -(\tilde c_{ql}^{(1)})^{\tau\tau}
\,,
\\
(\tilde c_{de})^{\mu\mu}&=(\tilde c_{dl})^{\mu\mu} = -(\tilde c_{de})^{\tau\tau}= -(\tilde c_{dl})^{\tau\tau}
\,.
\end{align}
Consequently, the enhancement of the muon neutrino contribution to $B\to K^{(*)}\nu\bar\nu$ is cancelled almost exactly by the suppression of the tau neutrino contribution 
and the final effect is unobservably small \cite{Altmannshofer:2014cfa}.

\subsection{331 Models}
In the so-called 331 models  based on the gauge group $SU(3)_C\times SU(3)_L\times U(1)_X$ FCNC processes receive tree-level contributions from  a new heavy neutral gauge boson $Z^\prime$ and through  $Z-Z^\prime$ mixing also from tree-level 
SM $Z$ boson exchanges. In this model, only left-handed quark currents are 
present so that $C_R$, $C_9^\prime$, and  $C_{10}^\prime$ vanish. Moreover, 
$\widetilde{c}_{ql}^{(3)}=0$. For the remaining coefficients of the 
effective theory, we find:
\be
\sin^2\theta_W\widetilde{c}_{ql}^{(1)}=-\left[\frac{\Delta_{L}^{\nu\nu}(Z')}{g^2_\text{SM}M_{Z'}^2}\right]
\frac{\Delta_{L}^{qb}(Z')}{ V_{tq}^\ast V_{tb}}\,,
\ee
\be 
\widetilde{c}_Z=R^L_{\nu\nu}\widetilde{c}_{ql}^{(1)}\,,
\ee
\be
\sin^2\theta_W\widetilde{c}_{qe}=-\left[\frac{\Delta_{R}^{\mu\mu}(Z')}{g^2_\text{SM}M_{Z'}^2}\right]
\frac{\Delta_{L}^{qb}(Z')}{ V_{tq}^\ast V_{tb}} \,.
\ee
where $\Delta_{L,R}^{ij}(Z^\prime)$ are the couplings defined in \cite{Buras:2014yna},
\be\label{gsm}
g_{\text{SM}}^2=4\frac{G_F}{\sqrt 2}\frac{\alpha}{2\pi\sin^2\theta_W}=1.78137\times 10^{-7} \gev^{-2}\,,
\ee
 and 
\be\label{DF1}
R^L_{\nu\nu}=\sin\xi \left[\frac{M_{Z\prime}^2}{M_{Z}^2}\right] \left[\frac{\Delta^{\nu\nu}_L(Z)}{\Delta^{\nu\nu}_L(Z^\prime)}\right]= B(\beta,a)\left[\frac{\Delta^{\nu\nu}_L(Z)}{\Delta^{\nu\nu}_L(Z^\prime)}\right].
\ee
Here $\sin\xi$  describes the $Z-Z^\prime$ mixing which depends 
 on two parameters $\beta$ and  $a = \left(1-\tan\bar\beta\right)/\left(1+\tan\bar\beta\right)$ and $M_{Z^\prime}$.  An explicit 
formula for $\sin\xi$ can be found in \cite{Buras:2014yna}.

The remarkable property of the  formula (\ref{DF1}) for $R^L_{\nu\nu}$
is its independence of $M_{Z^\prime}$ and this therefore also applies 
to the correlation between the coefficients $\widetilde{c}_Z$ and $\widetilde{c}_{ql}^{(1)}$, even if the range of the values of these coefficients depends 
on $M_{Z^\prime}$. This correlation depends on $\beta$ 
and  $\tan\bar\beta$ and on the fermion representations through $Z^\prime$ 
couplings. In \cite{Buras:2014yna}, 24 versions of 331 models 
have been considered, characterized by four values of $\beta$, three values of $\tan\bar\beta=0.2,~1.0,~5.0$,  and two fermion 
representations  $F_1$ and $F_2$. Among these 24 
possibilities, seven are favoured by electroweak precision tests. 

\begin{figure}[!tb]
\centering
\includegraphics[width=0.32\textwidth]{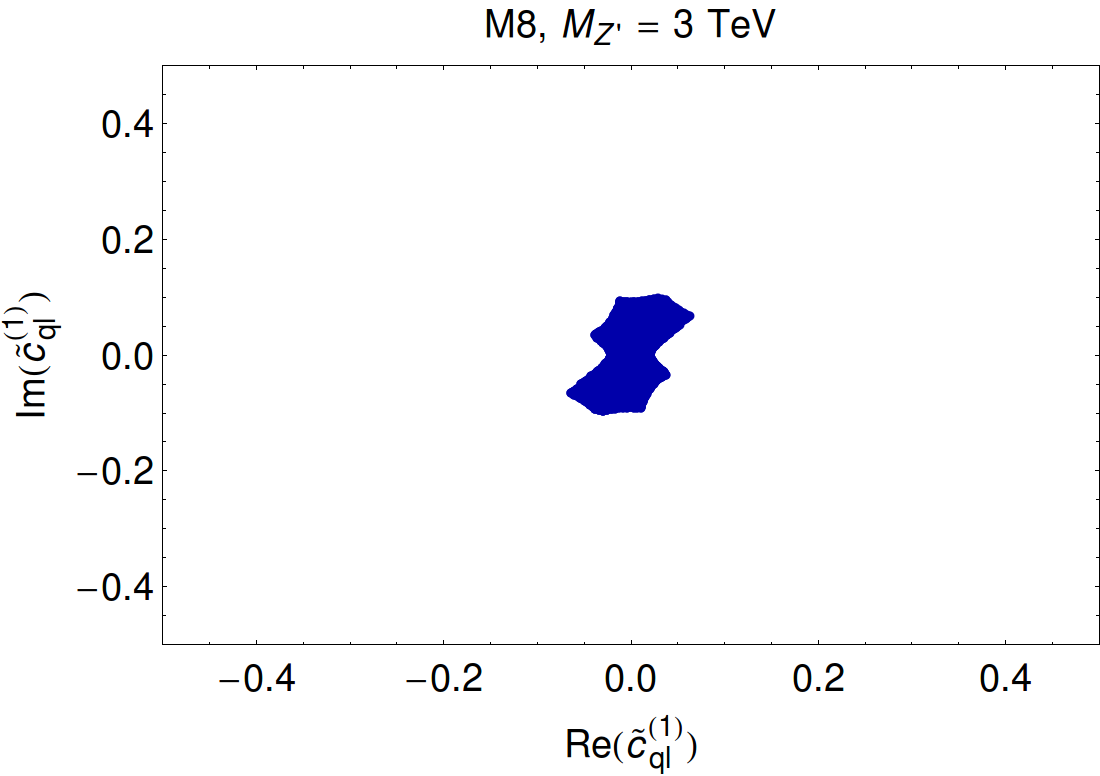}
\includegraphics[width=0.32\textwidth]{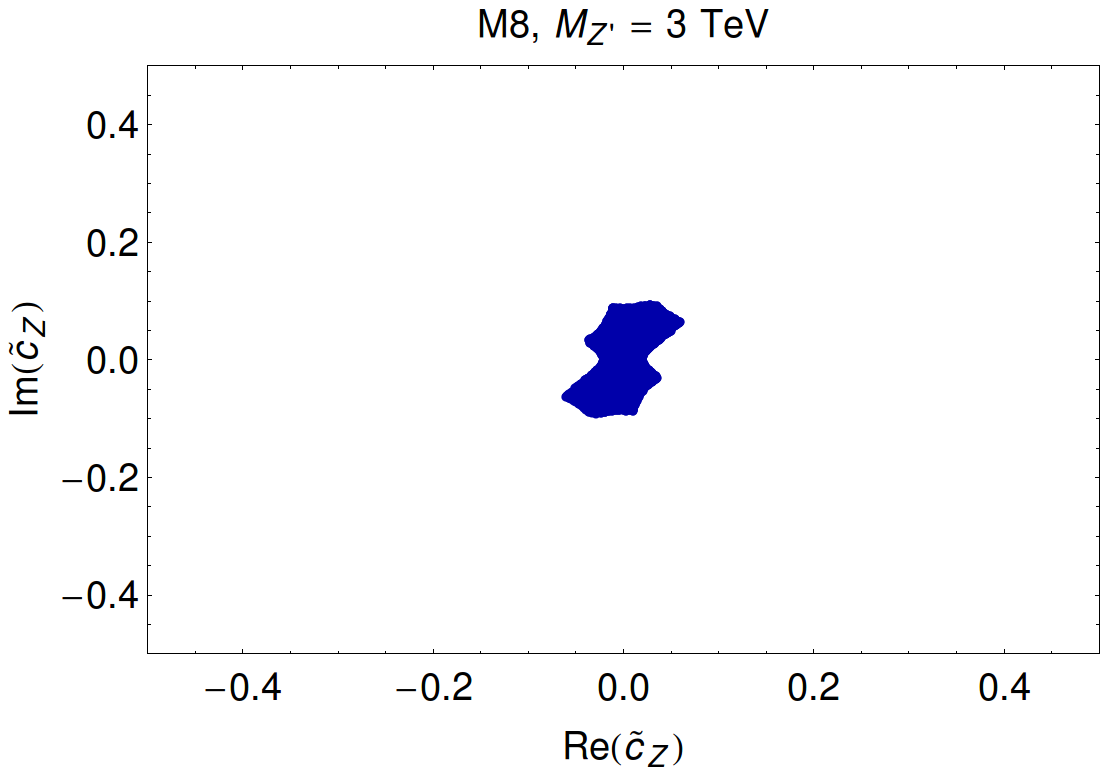}
\includegraphics[width=0.32\textwidth]{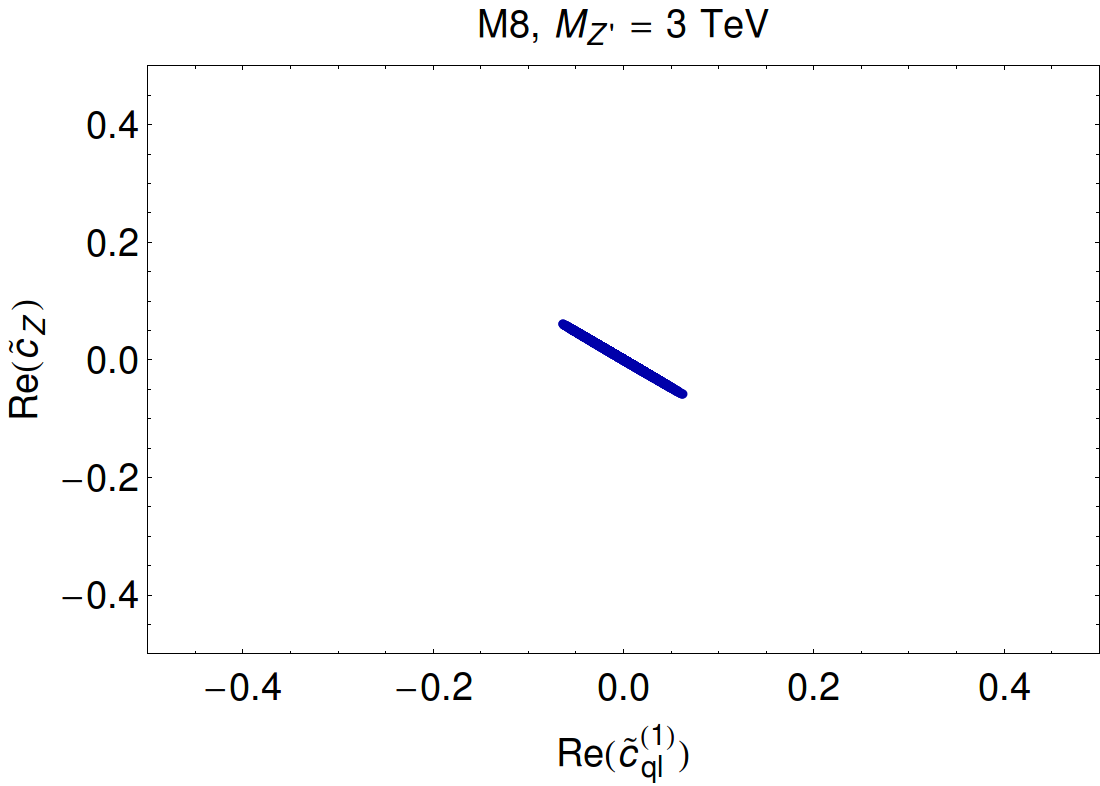}
\includegraphics[width=0.32\textwidth]{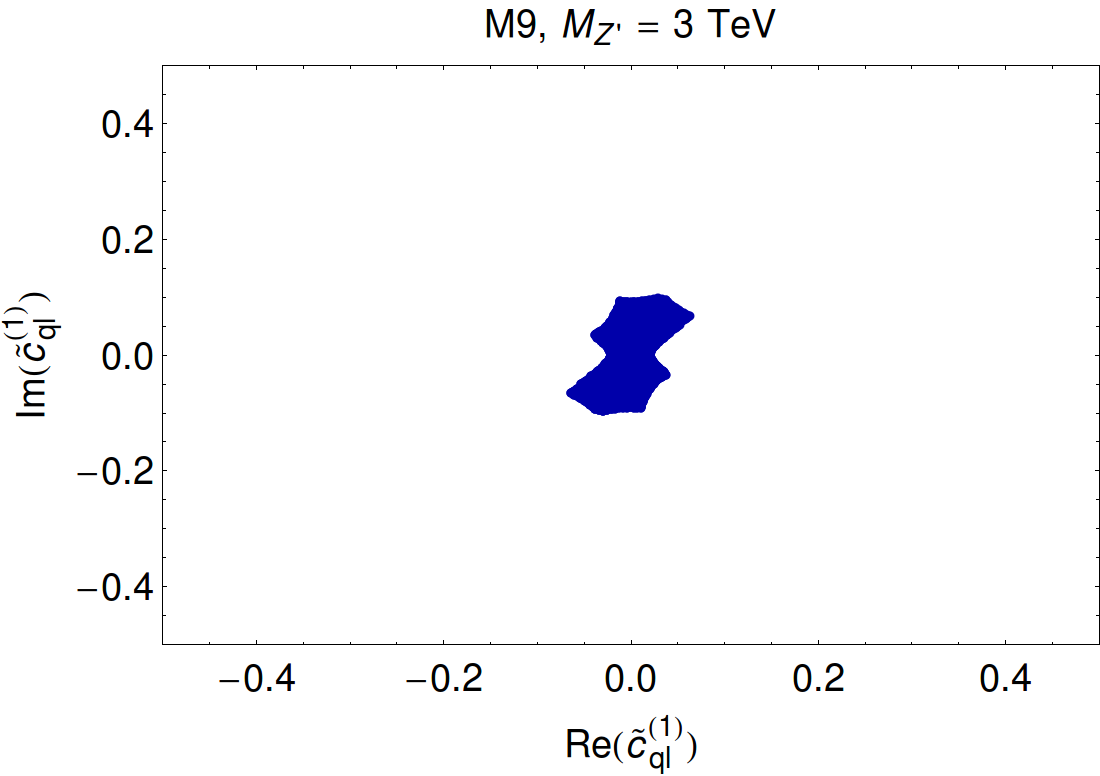}
\includegraphics[width=0.32\textwidth]{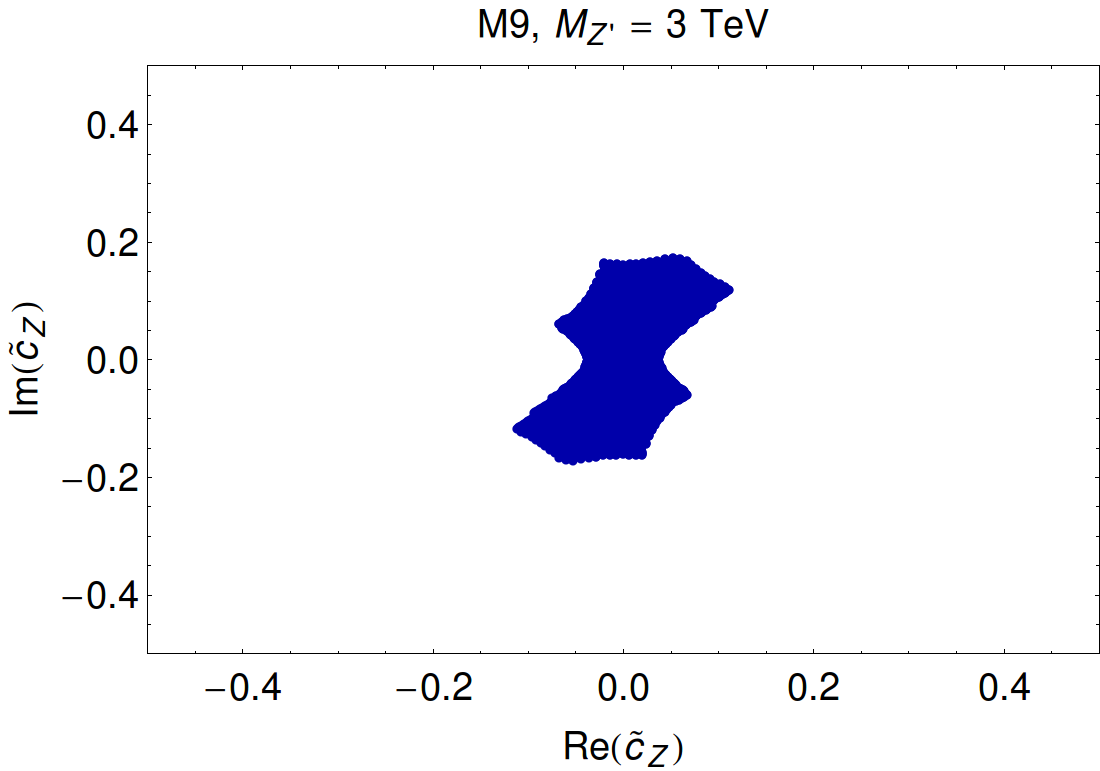}
\includegraphics[width=0.32\textwidth]{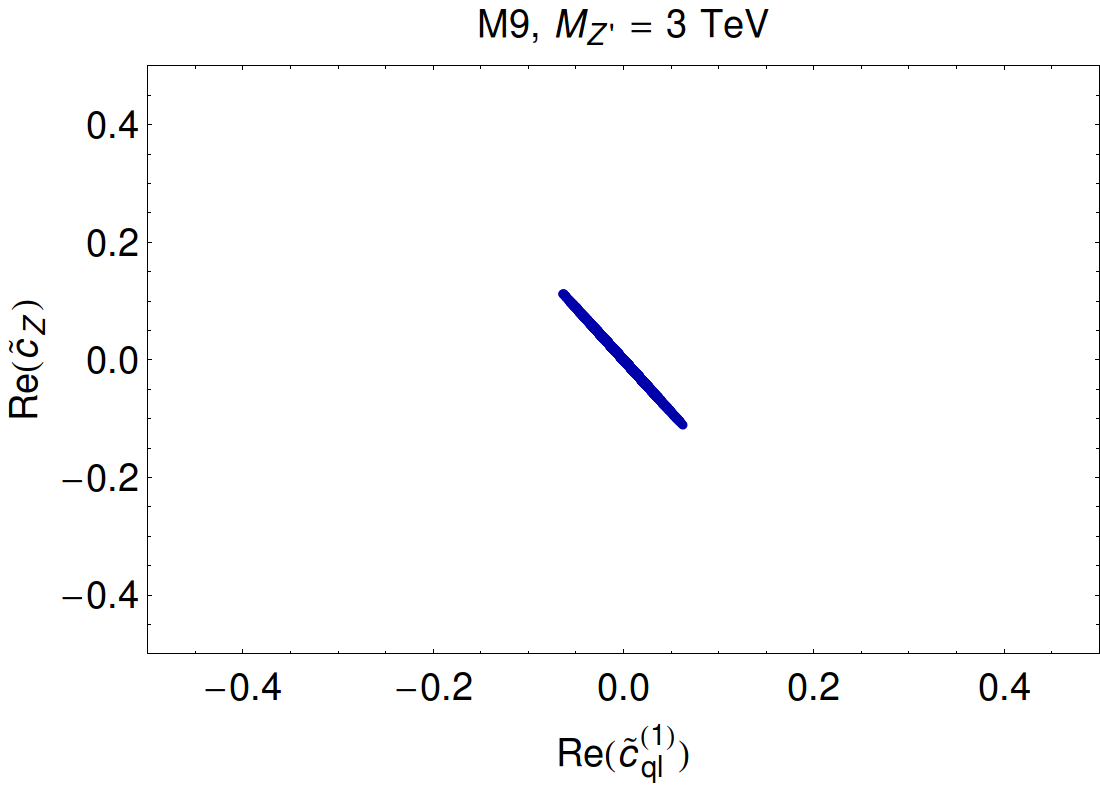}
\includegraphics[width=0.32\textwidth]{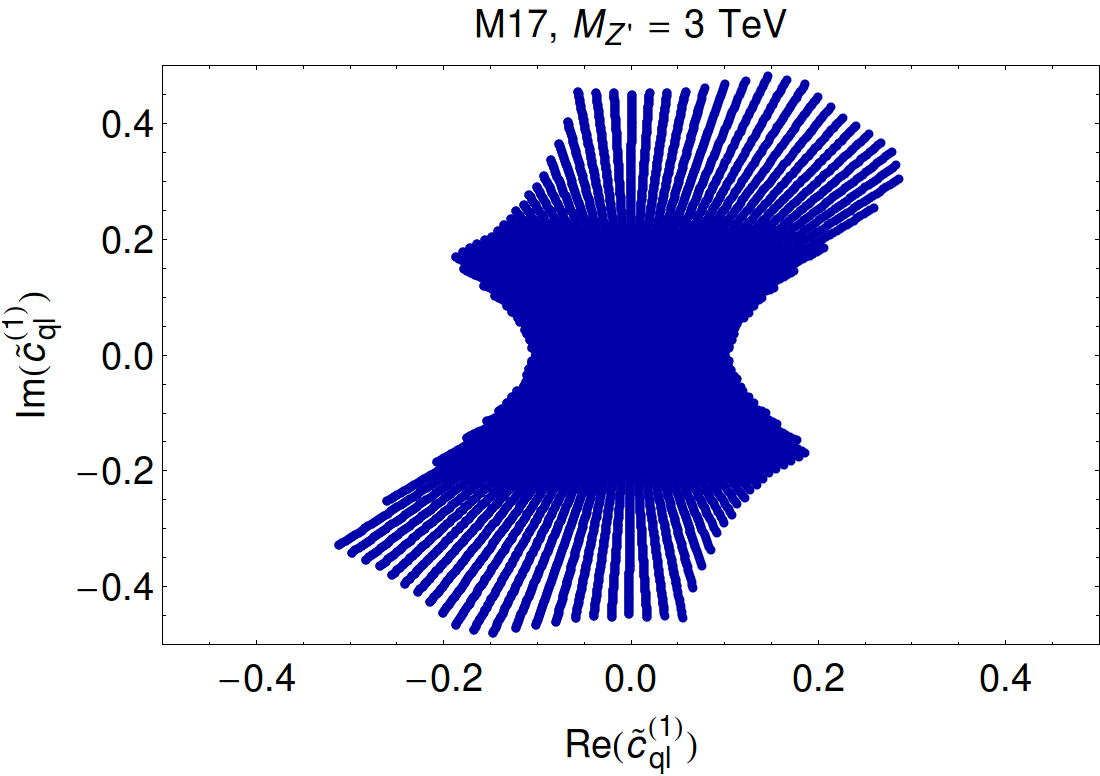}
\includegraphics[width=0.32\textwidth]{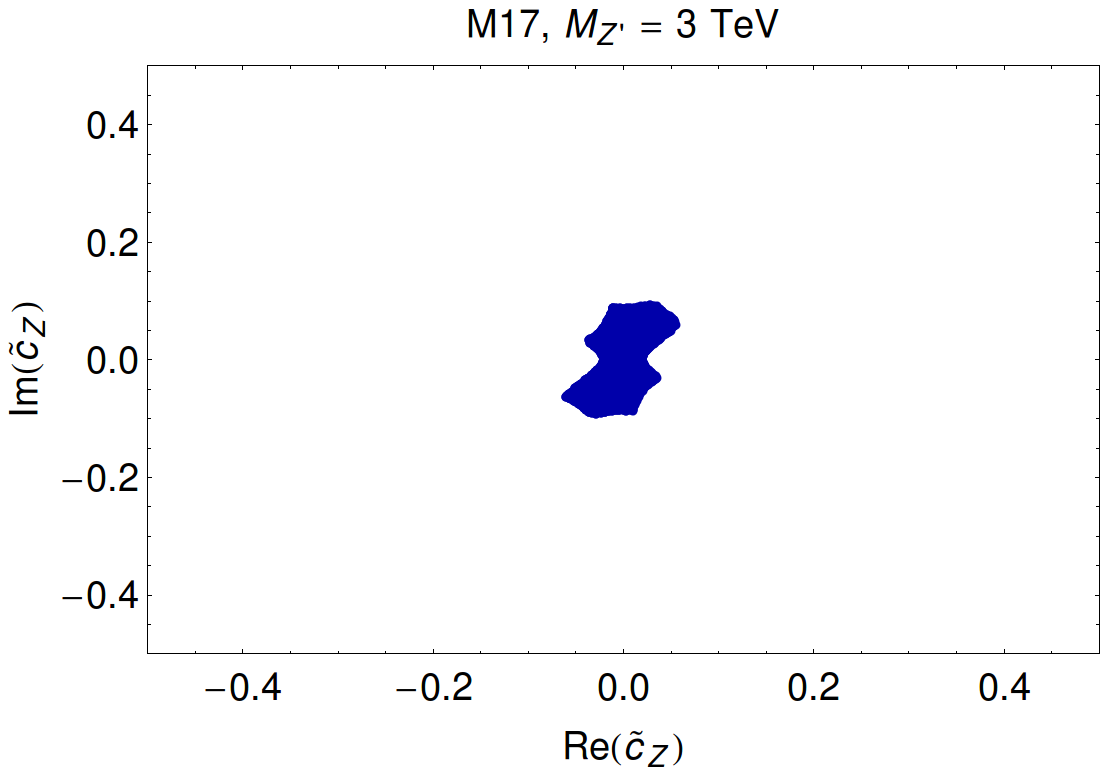}
\includegraphics[width=0.32\textwidth]{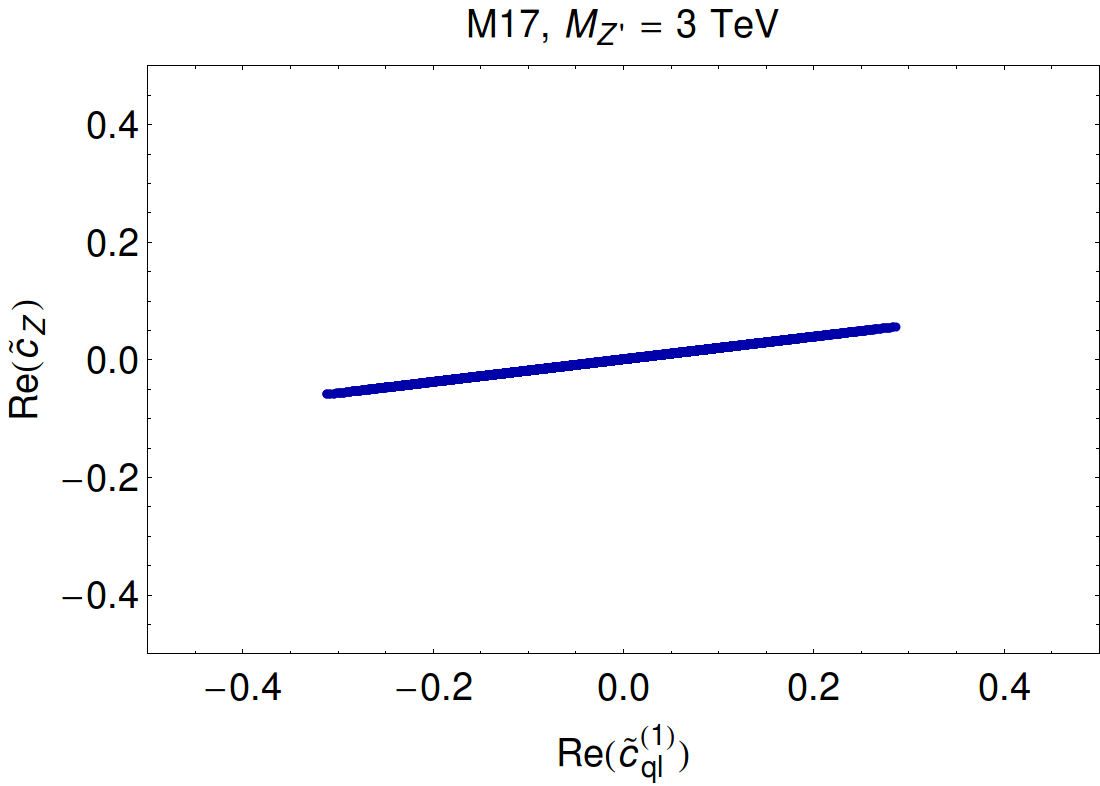}
\caption{Correlation between the Wilson coefficients $\widetilde{c}_Z$ and $\widetilde{c}_{ql}^{(1)}$ in three concrete versions of the 331 model for $M_{Z^\prime} = 3~$TeV
(M8: fermion representation F1, $\beta = 2/\sqrt{3}$, $\tan\bar\beta = 5$, M9: fermion representation F2, $\beta = -2/\sqrt{3}$, $\tan\bar\beta = 1$, 
M17: fermion representation F1, $\beta = -2/\sqrt{3}$, $\tan\bar\beta = 0.2$). Constraints from $b\to s\mu^+\mu^-$ transitions are included ($2\sigma$ range).}
\label{fig:wilsonM14}
\end{figure}

In figure~\ref{fig:wilsonM14}, we show exemplarily the correlation between the coefficients $\widetilde{c}_Z$ and $\widetilde{c}_{ql}^{(1)}$ for three cases
(M8, M9 and M17) that are defined in the caption of this figure. M8 and M9 are the leaders among the seven models that pass electroweak precision tests. M17 
can only be accepted if the LEPII result for the asymmetry $A_l$ is declared to be correct and the SLD result ignored. We show this case as it features 
different pattern of flavour violation than M8 and M9 cases. The following comments should be made.
\begin{itemize}
 \item 
 The results of the other  favoured models look similar in shape to the ones for M8 and M9 but differ a bit
in magnitude. In particular in all seven favoured cases there is an anticorrelation between 
$\widetilde{c}_Z$ and $\widetilde{c}_{ql}^{(1)}$ implying significant 
cancellation between $Z^\prime$ and $Z$ contribution to $b\to s\nu\bar\nu$ 
channels. This cancellation can only be seen in a concrete model and cannot 
be predicted within an effective theory approach. 
\item
As NP contributions to $C_{10}$ are governed in these models by the term $\widetilde{c}_Z-\widetilde{c}_{ql}^{(1)}$, the anticorrelation between these two coefficients implies 
 constructive interference between $Z^\prime$ and $Z$ contributions to $B_s\to\mu^+\mu^-$. This means for instance that $Z$ and $Z^\prime$ can jointly suppress the 
 rate of $B_s\to\mu^+\mu^-$ as appears to be required by the data.
\item
In the case of M17, $\widetilde{c}_Z$ and $\widetilde{c}_{ql}^{(1)}$
are however correlated implying larger contributions to $b\to s\nu\bar\nu$ 
channels but enhancing the rate of $B_s\to\mu^+\mu^-$ which is disfavoured by the data.
\end{itemize}

Due to the absence of right-handed currents, an important prediction of these models is 
\begin{equation}
\mathcal{R}_K=\mathcal{R}_K^{*}\,.
\end{equation}
As this relation is also valid in 
models with MFV, 331 models having 
new sources of flavour and CP violation can be best distinguished from MFV models through CP-violating quantities and other correlations presented in \cite{Buras:2014yna}.

One can also understand the pattern of NP effects in these models in terms of the SM-EFT, as 
confirmed in \cite{Buras:2014yna}:
\begin{itemize}
\item
$Z$ contributions to $C_9$ are strongly suppressed because of $\zeta$ being 
small.
\item 
The  fact that the 
sum $\widetilde{c}_{qe}+\widetilde{c}_{ql}^{(1)}$ enters $C_9$, while 
the difference $\widetilde{c}_{qe}-\widetilde{c}_{ql}^{(1)}$ enters 
$C_{10}$, shows that it is not easy to get simultaneously significant NP 
contributions to $B\to K^*\mu^+\mu^-$ and $B_s\to\mu^+\mu^-$. This 
is also found in numerous plots in \cite{Buras:2014yna}.
\end{itemize}

\begin{figure}
\centering
\includegraphics[width=0.45\textwidth]{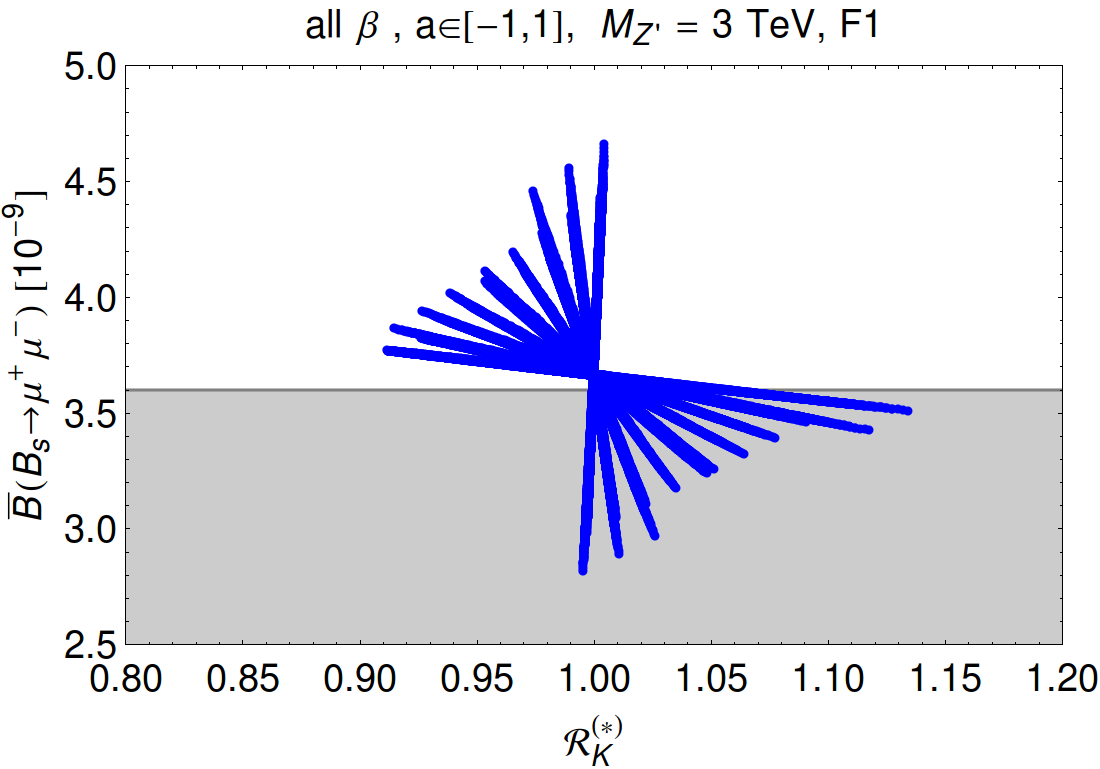}
\includegraphics[width=0.45\textwidth]{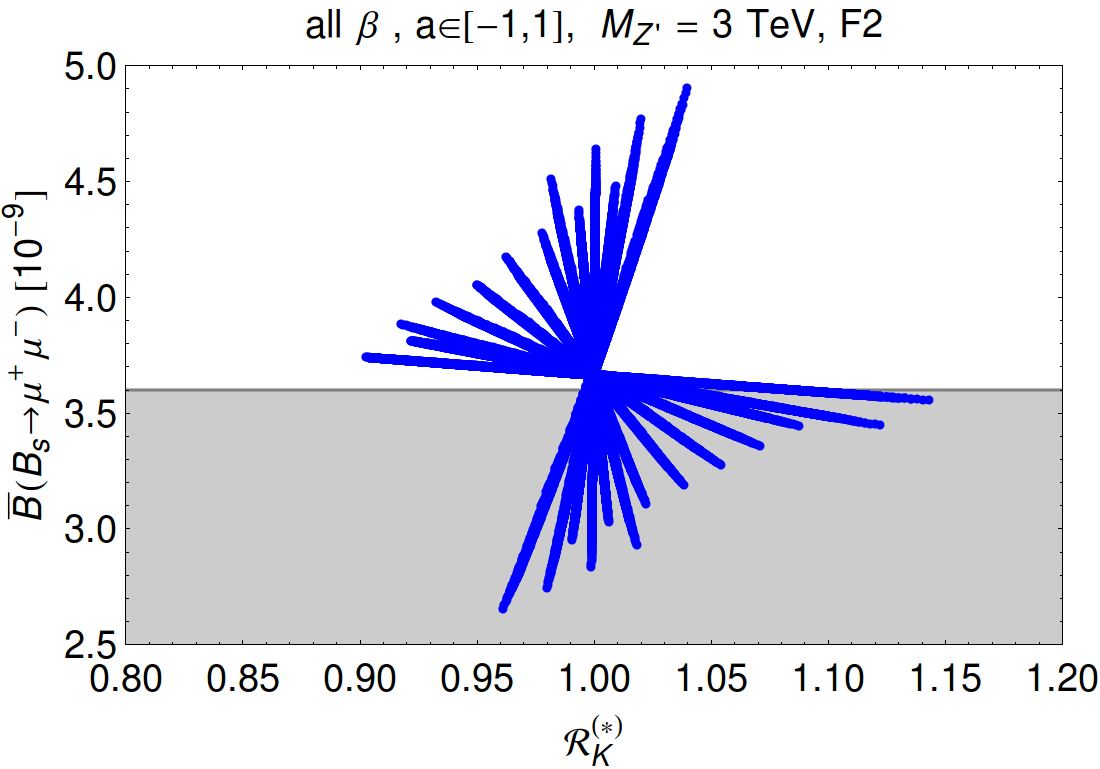}
\caption{Correlation
 $\overline{\text{BR}}(B_s\to\mu^+\mu^-)$ versus $\mathcal{R}_K^{(*)}$ in the 331 model for $M_{Z^\prime} = 3~$TeV for fermion representations F1 (left) and F2 (right).}
\label{fig:BsmuvsBK}
\end{figure}

In figure~\ref{fig:BsmuvsBK}, we show the correlation between  $\overline{\text{BR}}(B_s\to\mu^+\mu^-)$ and $\mathcal{R}_K^{(*)}$ for 
both fermion representations. In contrast to \cite{Buras:2014yna}, where we considered in total 24 versions of the model, we study here
an even larger set, since we do not fix the model parameter $a$ to three different values. 
Here we set $M_{Z^\prime} = 3~$TeV, scanned over  $a\in[-1,1]$ (corresponding to $\tan\bar\beta = [0,\infty]$) and over $\beta = \pm 
2/\sqrt{3},\pm1/\sqrt{3}$.
Constraints from $b\to s \ell^+\ell^-$ transitions ($2\sigma$ range) and electroweak observables as in~\cite{Buras:2014yna} ($\Omega^{331}\leq 16$) are included.
One can see that even when combining all models a suppression of $B_s\to\mu^+\mu^-$, as favoured by present data, 
almost always implies an enhancement of $b\to s\nu\bar\nu$. Models where both are enhanced or suppressed
simultaneously are excluded due to electroweak observables constraints.

On the whole, 331 models are an example that specific NP models can be much more predictive than a generic EFT approach. 
The size of the NP effects in $b\to s\nu\bar\nu$ in 331 models turn out 
to be small, typically below $15\%$ at the level of the branching ratios.

\subsection{Partial Compositeness}

\begin{figure}[tbp]
\centering
\includegraphics[width=0.4\textwidth]{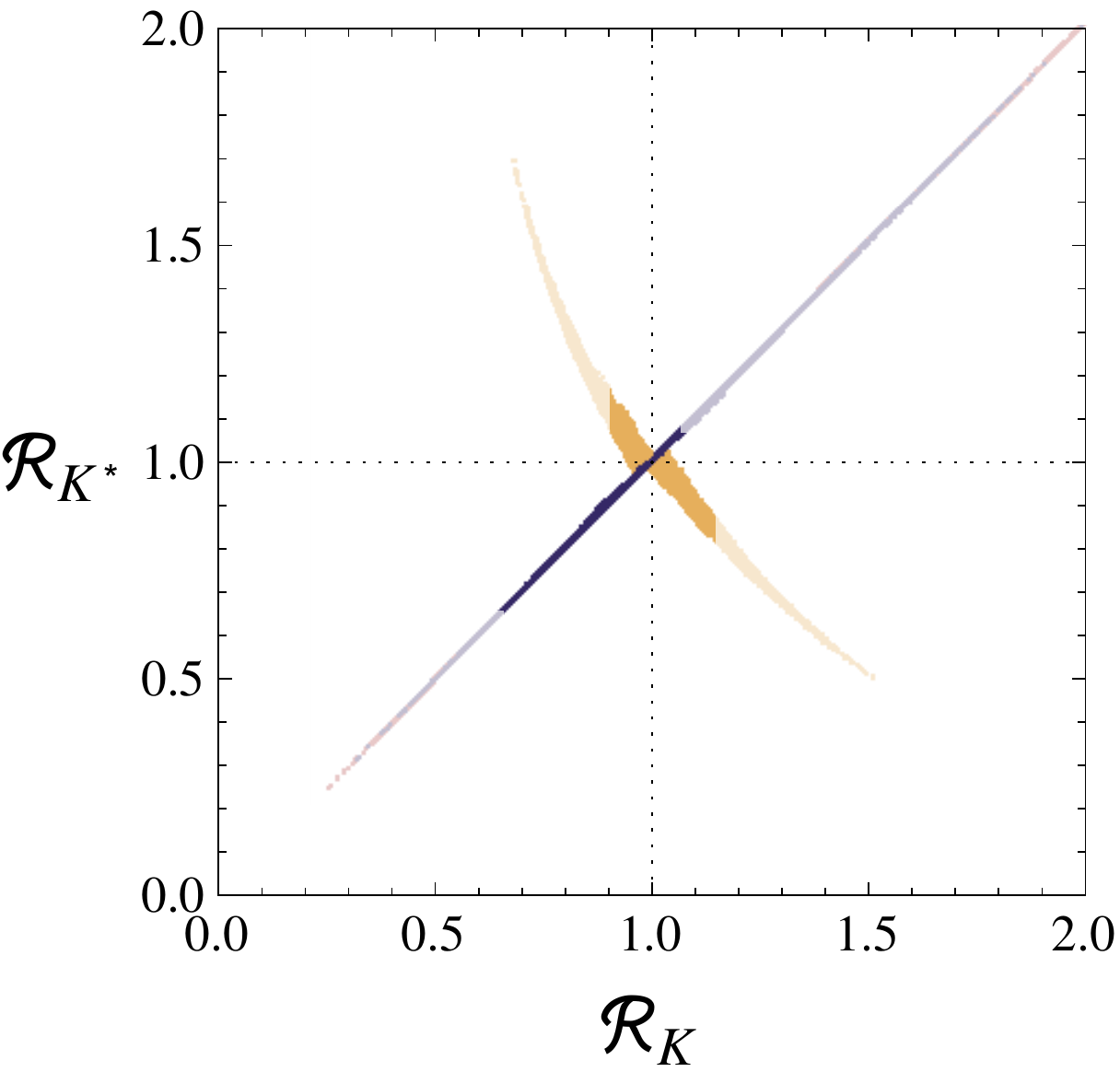}
\caption{Allowed ranges for the $B\to K\nu\bar\nu$ and $B\to K^{*}\nu\bar\nu$ branching ratios in models with partial compositeness, normalized to the SM values, for two different choices of fermion representations: bidoublet model (blue),  triplet model (yellow). Light points are disfavoured at $2\sigma$ by $b\to s\mu^+\mu^-$ data. Plot adapted from \cite{Straub:2013zca}.
}
\label{fig:PC}
\end{figure}

Partial quark compositeness is a feature of composite Higgs models and of the four-dimensional Kaluza-Klein picture of models with  extra dimensions. 
Rare $B$ decays in a simple 4D partial compositeness model with different choices for the flavour structure and the representations of composite fermions have 
been considered in \cite{Straub:2013zca}. The dominant contributions to $b\to s\nu\bar\nu$ transitions in these models come from tree-level 
flavour-changing $Z$ couplings, i.e. contributions to the operators $\widetilde{c}_Z$ (in the ``bidoublet model'') or $\widetilde{c}_Z'$ (in the ``triplet model''). 
Consequently, the bounds in (\ref{eq:boundcZ}) apply and limit the size of NP effects in $B\to K^{(*)}\nu\bar\nu$.
The accessible range for $\mathcal R_K$ and $\mathcal R_{K^*}$ is shown in figure~\ref{fig:PC}. The blue points correspond to the bidoublet model, the yellow points to the triplet model. The lighter points are disfavoured at $2\sigma$ by $b\to s\mu^+\mu^-$ data.

This can be compared to the Randall-Sundrum model with custodial protection (RSc) studied in \cite{Blanke:2008yr,Biancofiore:2014uba}. The fermion representations chosen in this model are 
similar to the triplet model of \cite{Straub:2013zca} and the NP effects in $b\to \nu\bar\nu$ are dominated by $\widetilde{c}_Z'$. 
However, since the extra dimensional model is much more restrictive, linking e.g. the scale of fermion resonances to that of the vector resonances, which are more strongly constrained experimentally, the maximal allowed effects in the RSc are significantly smaller.
 Larger effects were found in the Randall-Sundrum model without custodial protection \cite{Bauer:2009cf}, however in that case it is difficult to fulfill 
 electroweak precision constraints, in particular from the $T$ parameter.

Effects in Wilson coefficients other than $\tilde c_{Z}$ and  $\tilde c_{Z}'$ could be generated in models with partial compositeness by heavy vector resonance exchange if any of the leptons has a sizable degree of compositeness for its left-handed chirality (the products of degrees of compositeness of both chiralities have to be small because of the leptons' lightness). Depending on the representations of the composite fermions, one can then generate contributions to the Wilson coefficients $(\tilde c_{ql}^{(1)})^{\ell\ell}$, $(\tilde c_{ql}^{(3)})^{\ell\ell}$, and $(\tilde c_{dl})^{\ell\ell}$, where $\ell=e,\mu$, or $\tau$. Up to model-dependent (complex) $O(1)$ factors, the Wilson coefficients are parametrically given by
\begin{align}
(\tilde c_{ql}^{(1,3)})^{\ell\ell}
&\sim
\frac{g_\rho^2}{V_{tb}V_{ts}^*} s_{Lb}s_{Ls}s_{L\ell}^2
\left[\frac{5\,\text{TeV}}{m_\rho}\right]^2
\,,
&
(\tilde c_{dl})^{\ell\ell}
&\sim
\frac{g_\rho^2}{V_{tb}V_{ts}^*} s_{Rb}s_{Rs}s_{L\ell}^2
\left[\frac{5\,\text{TeV}}{m_\rho}\right]^2
\,,
\end{align}
where $g_\rho$ and $m_\rho$ are a typical coupling and mass scale of the vector resonances and $s_{Lf,Rf}$ is the degree of compositeness of the left-handed or right-handed fermion.
As an example, we can consider models with flavour anarchy, where one expects $s_{Lb}s_{Ls}\sim V_{tb}V_{ts}^*\sim0.04$ and  $s_{Rb}s_{Rs}\sim m_dm_s/v^2/(V_{tb}V_{ts}^*)\sim0.01$. For $m_\rho/g_\rho\sim 1\,$TeV, we see that, barring an additional enhancement, visible effects in the 4-fermion operators require an $O(1)$ degree of lepton compositeness.
In general, one then also expects corrections to $Z\ell_L\ell_L$ couplings of order $s_{L\ell}^2g_\rho^2/m_\rho^2$ that are excluded by LEP precision measurements at the $Z$ pole for $s_{L\ell}$ of $O(1)$. However, in models where the $Z\ell_L\ell_L$ couplings are protected by a custodial symmetry (see e.g.~\cite{Agashe:2009tu}), such scenario could still be viable. An exhaustive analysis of this scenario is beyond the scope of our analysis.

\subsection{MSSM}

In the MSSM, the dominant NP effects in $b\to s\nu\bar\nu$ arise from $Z$ penguins, i.e. through $\widetilde{c}_Z$ and $\widetilde{c}_Z'$. While the former can be 
generated in the MSSM with MFV, the latter requires non-minimal flavour violation. However, it has been shown already in \cite{Altmannshofer:2009ma} that 
$\widetilde{c}_Z'$ is very small throughout the MSSM parameter space once constraints from other flavour observables (notably $B_s\to\mu^+\mu^-$) are taken 
into account and that sizable effects in $\widetilde{c}_Z$ are only possible beyond MFV, in particular in the presence of a flavour-changing trilinear coupling in the up-type squark sector.

In view of the improved constraints on both $\Delta F=1$ and $\Delta F=2$ observables in the $b\to s$ sector as well as improved direct bounds on sparticle masses, we have performed a numerical analysis of the MSSM parameter space to asses the maximal size of NP effects in $b\to s\nu\bar\nu$ still allowed in the MSSM. Our starting point is the 24-parameter phenomenological MSSM, to which we add all off-diagonal terms in the squark mass matrices and trilinear couplings relevant for $b\to s$ transitions. The flavour diagonal parameters are scanned in the following ranges,
\begin{align}
M_1 &\in [1,1500] \,\text{GeV}, &
m_{\tilde Q_{1}},m_{\tilde U_{1}},m_{\tilde D_{1}} &\in [400,3000] \,\text{GeV}, \\
M_2 &\in [100,1500] \,\text{GeV}, & 
m_{\tilde Q_{3}},m_{\tilde U_{3}},m_{\tilde D_{3}} &\in [400,3000] \,\text{GeV}, \\
M_3 &\in [400,3000] \,\text{GeV}, &
m_{\tilde L_{1}},m_{\tilde \nu_{1}} &\in [100,3000] \,\text{GeV}, \\
|\mu| &\in [100,1500] \,\text{GeV}, &
m_{\tilde L_{3}},m_{\tilde \nu_{3}} &\in [100,1500] \,\text{GeV}, \\
M_A &\in [100,1500] \,\text{GeV} & A_{u,d,l} &\in [-3000,3000]\,\text{GeV},
\end{align}
where the trilinear parameters are scanned linearly and all others logarithmically. The gaugino masses and the $\mu$ term are assumed to be real; both signs are allowed for $\mu$.
We define the mass insertions as
\begin{align}
(\delta^{LL})_{ij} &= \frac{(m_Q^2)_{ij}}{\sqrt{(m_Q^2)_{ii}(m_Q^2)_{jj}}} \,,
&
(\delta^{RR}_{u,d})_{ij} &= \frac{(m_{U,D}^2)_{ij}}{\sqrt{(m_{U,D}^2)_{ii}(m_{U,D}^2)_{jj}}} \,,
\\
(\delta^{LR}_{u,d})_{ij} &= \frac{(T_{U,D})_{ij}}{\sqrt{(m_Q^2)_{ii}(m_{U,D}^2)_{jj}}} \,,
\end{align}
and scan $|(\delta^{LR}_{u,d})_{23,32}|$, $|(\delta^{RR}_{u,d})_{23}|$, and $|(\delta^{LL})_{23}|$ logarithmically between $10^{-4}$ and 1, allowing for an arbitrary phase.
Finally, we require the lightest neutralino to be the LSP and use \texttt{FastLim 1.0} \cite{Papucci:2014rja} to impose LHC bounds on sparticle masses and \texttt{SUSY\char`_FLAVOR 2.11} \cite{Crivellin:2012jv} to compute the $b\to s\nu\bar\nu$ Wilson coefficients and impose FCNC constraints, in particular BR($B\to X_s\gamma$), BR($B_s\to\mu^+\mu^-$), $\Delta M_s$, and $\phi_s$.

\begin{figure}
\centering
\includegraphics[width=0.45\textwidth]{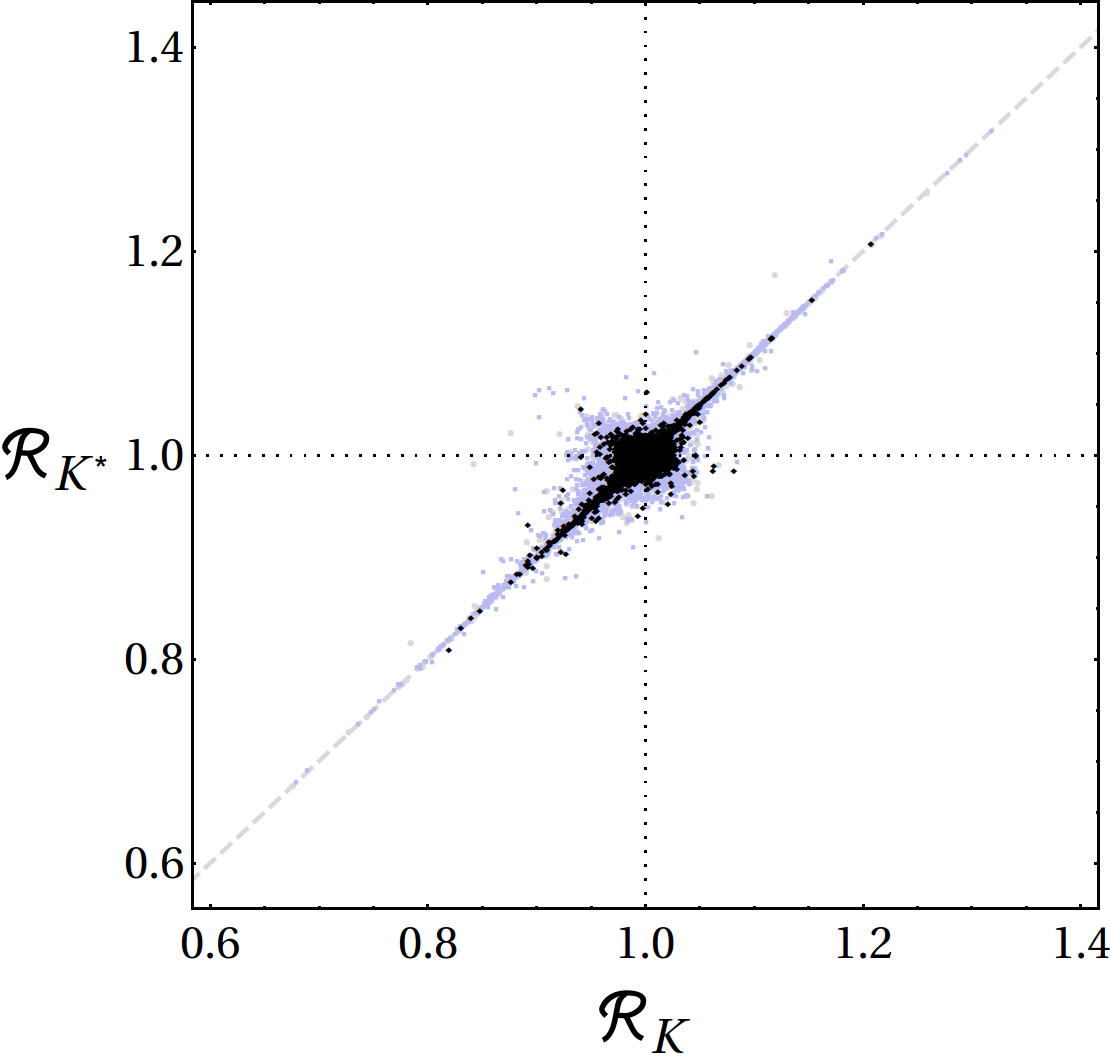}
\caption{Allowed ranges for the $B\to K\nu\bar\nu$ and $B\to K^{(*)}\nu\bar\nu$ branching ratios in the MSSM, normalized to the SM values. All dark points pass flavour and collider constraints; black points have the corrected lightest Higgs mass.}
\label{fig:RK-SUSY}
\end{figure}

The result of the parameter scan is shown in figure~\ref{fig:RK-SUSY}. All points shown are allowed by flavour constraints. Light gray points are ruled out by direct LHC bounds. Among the allowed points, we distinguish the black ones that have a lightest Higgs mass (computed with \texttt{SPheno 3.3.2} \cite{Porod:2003um,Porod:2011nf}) within 4~GeV of the true value of 125~GeV from the light blue ones, that have a too light or too heavy lightest Higgs.
Our rationale for showing these points as well is that they might be realized in extensions of the MSSM raising the tree-level Higgs mass.

The numerical results confirm the findings of \cite{Altmannshofer:2009ma} that right-handed current contributions to $b\to s\nu\bar\nu$ are small in the MSSM, so the relation $\mathcal{R}_K=\mathcal{R}_{K^*}$, indicated by a dashed line in figure~\ref{fig:RK-SUSY}, is approximately fulfilled.
The $B\to K\nu\bar\nu$ and $B\to K^{(*)}\nu\bar\nu$ can be enhanced or suppressed by at most 30\% relative to the SM. This conclusion is not changed by the existing LHC direct bounds on sparticle masses, which have been taken into account in our scan as described above.

\subsection{Leptoquarks}\label{sec:lq}

In models with leptoquarks one assumes the presence of heavy (scalar or vector) particles which carry colour and, thus, lead to interactions connecting leptons and quarks. This generically happens in GUTs or in SUSY theories with R-parity violation. As a consequence, the four-fermion operators (\ref{eq:ops}) relevant for $b \rightarrow s \nu \bar \nu$ and $b \rightarrow s \ell^+ \ell^-$ processes  can be generated by a tree-level exchange of the heavy leptoquark.

\begin{table}
\renewcommand{\arraystretch}{2.3}
\centering
 \begin{tabular}{ccccr}
 \hline
& Spin & $G_\mathrm{SM}$  & interaction term & generated Wilson coefficients \\
 \hline
 $S_1$ & 0 & $\left(\overline{\mathbf{3}}, \mathbf{1} \right)_{\frac{1}{3}}$ & $\lambda_{i j} \, \left( \overline{q^c_{{L} i}} \cdot \epsilon \cdot l_{{L} j } \right) \, \phi$ & $\frac{\left[c_{q l}^{(1)}\right]_{ij;kl}}{\Lambda^2} = - \frac{\left[c_{q l}^{(3)}\right]_{ij;kl}}{\Lambda^2} = - \frac{1}{4}\frac{\lambda_{j l} \lambda^*_{i k}}{m_\phi^2}$\\
 $S_3$ & 0 & $\left( \overline{\mathbf{3}}, \mathbf{3} \right)_{\frac{1}{3}}$ & $\lambda_{i j} \, \left( \overline{q^c_{{L} i}} \cdot \epsilon \cdot \tau^a \cdot l_{{L} j} \right) \, \phi^a$ & $\frac{\left[c_{q l}^{(1)} \right]_{i j ; k l}}{\Lambda^2} =  3\frac{\left[c_{q l}^{(3)}\right]_{i j ; k l}}{\Lambda^2} = \frac{3}{4} \frac{\lambda_{j l} \lambda^*_{i k}}{m_\phi^2}$ \\
 $\widetilde{R}_2$ & 0 & $\left( \mathbf{3}, \mathbf{2} \right)_{\frac{1}{6}}$ & $\lambda_{i j} \, \overline{d_{{R}i}} \left( l_{{L}j} \cdot \epsilon \cdot \phi \right)$ & $\frac{\left[ c_{d l} \right]_{i j ; k l}}{\Lambda^2} = -\frac{1}{2} \frac{\lambda_{i l} \lambda^*_{j k}}{m_\phi^2}$ \\
 $U_1$ & 1 & $\left( \mathbf{3}, \mathbf{1} \right)_{\frac{2}{3}}$ & $\lambda_{i j} \, \left( \overline{q_{{L}i}} \, \gamma^\mu \, l_{{L}j} \right) \, \phi_\mu$ & $\frac{\left[c_{q l}^{(1)}\right]_{i j ; k l}}{\Lambda^2}  = \frac{\left[c_{q l}^{(3)}\right]_{i j ; k l}}{\Lambda^2} = \frac{1}{2}\frac{\lambda_{i l} \lambda^*_{j k}}{m_\phi^2}$ \\
 $U_3$ & 1 & $\left( \mathbf{3}, \mathbf{3} \right)_{\frac{2}{3}}$ & $\lambda_{i j} \, \left( \overline{q_{{L}i}} \, \gamma^\mu \, \tau^a \, l_{{L}j} \right) \, \phi^a_\mu$ & $\frac{\left[c_{q l}^{(1)}\right]_{i j; k l}}{\Lambda^2} = - 3 \frac{\left[c_{q l}^{(3)}\right]_{i j ; k l}}{\Lambda^2} = - \frac{3}{2}\frac{\lambda_{i l} \lambda^*_{j k}}{m_\phi^2}$ \\ 
 $V_2$ & 1 & $\left( \overline{\mathbf{3}}, \mathbf{2} \right)_{\frac{5}{6}}$ & $\lambda_{i j} \, \overline{d_{{R}i}^c} \, \gamma^\mu \left( l_{{L}j} \cdot \epsilon \cdot \phi_\mu \right)$ & $\frac{\left[ c_{d l} \right]_{i j ; k l}}{\Lambda^2} = \frac{\lambda_{i l} \lambda^*_{j k}}{m_\phi^2}$\\
 \hline
 \end{tabular}
\caption{Possible leptoquark scenarios relevant for $b \rightarrow s \nu \bar{\nu}$ decays. In the first columns, the spin and gauge quantum numbers are given as well as the relevant interaction term. In the last column, we give expressions for the Wilson coefficients of the generated four-fermion operators. The SM left-handed quark and lepton doublets are denoted by $Q_{L}$ and $L_{L}$, respectively, while the leptoquark is written as $\phi_{(\mu)}$. We explicitly showed the flavour indices here.}
\label{tab:lp}
\end{table}

The number of leptoquark models is strongly restricted by the assumption of SM gauge invariance. An extensive investigation of the viable scenarios is given in \cite{Buchmuller:1986zs}. However, not all possible scenarios lead to $b \rightarrow s \nu \bar \nu$ transitions. We summarize the viable options in Table \ref{tab:lp} and give expressions for the Wilson coefficients of the generated operators. Generally, these models are not lepton flavour universal and even flavour violating. For the time being we assume that the leptoquarks only couple to one lepton flavour. 

We find that the Wilson coefficients are strongly correlated. In models in which the leptoquark is an $\mathrm{SU}(2)$ singlet or triplet only the operators $Q_{ql}^{(1)}$ and $Q_{q l}^{(3)}$ are generated, but predicted to obey the relation
\begin{equation}
 \tilde c_{q l}^{(1)} = n \cdot \tilde c_{q l}^{(3)},
\end{equation}
where $n$ is some model-dependent real constant. From (\ref{eq:WC1})--(\ref{eq:WC3}) we then find that, for a given $n$, the low-energy Wilson coefficients only depend on one parameter,
\begin{align}
 C_L^\mathrm{NP} &= \tilde c_{q l}^{(1)} - \tilde c_{q l}^{(3)} = (n-1) \tilde c_{q l}^{(3)}, \\
 C_9^\mathrm{NP} = - C_{10}^\mathrm{NP} &= \tilde c_{q l}^{(1)} + \tilde c_{q l}^{(3)} = (n+1) \tilde c_{q l}^{(3)}, \\ 
 C_R = C_9' = C_{10}' &= 0.
\end{align}
Hence, we can write the corrections to the $b \rightarrow s \nu \bar \nu$ branching ratios in the following way,
\begin{equation}
 \mathcal{R}_{F_L} = 1, \qquad \mathcal{R}_K = \mathcal{R}_{K^*} = \frac{2}{3} + \frac{1}{3} \frac{\left| C_L^\mathrm{SM} + (n-1) \left[\tilde c_{q l}^{(3)}\right]_\ell \right|^2}{\left| C_L^\mathrm{SM} \right|^2}.
\end{equation}

In the case of a doublet leptoquark, only $Q_{d l}$ is present. So again, we expect only a dependence on one parameter,
\begin{align}
 C_L^\mathrm{NP} = C_9^\mathrm{NP} = C_{10}^\mathrm{NP} &= 0, \\ 
 C_R = C_9' = -C_{10}' &= \tilde c_{d l}.
\end{align}
In this case, we find a contribution to $\eta \neq 0$ such that also $\mathcal{R}_K \neq \mathcal{R}_{K^*}$,
\begin{align}
 \mathcal{R}_K &= \frac{2}{3} + \frac{1}{3} \left( 1 + 2 \frac{C_L^\mathrm{SM} \, \mathrm{Re}\left( \left[\tilde c_{d l}\right]_\ell \right)}{| C_L^\mathrm{SM} |^2 + |\left[\tilde c_{d l}\right]_\ell |^2} \right) \left( 1 + \frac{\left| \left[\tilde c_{d l}\right]_\ell \right|^2}{\left| C_L^\mathrm{SM} \right|^2} \right), \\
 \mathcal{R}_{K^*} &= \frac{2}{3} + \frac{1}{3} \left( 1 - \kappa_\eta \frac{C_L^\mathrm{SM} \, \mathrm{Re}\left( \left[\tilde c_{d l}\right]_\ell \right)}{| C_L^\mathrm{SM} |^2 + |\left[\tilde c_{d l}\right]_\ell |^2} \right) \left( 1 + \frac{\left| \left[\tilde c_{d l}\right]_\ell \right|^2}{\left| C_L^\mathrm{SM} \right|^2} \right).
\end{align}

From table \ref{tab:lp}, we can already see two special cases. In scenario $U_1$ there is $n=1$ which implies that all contributions to $\mathcal{R}_K=\mathcal{R}_{K^*}$ vanish such that we do not expect any deviation from the SM values in this model. In the scenario $S_1$, on the other hand, we find $n=-1$, which means that this scenario does not give any contribution to the decay into charged leptons. Hence, the effects in $\mathcal{R}_K$ and $\mathcal{R}_{K^*}$ are unconstrained from these decays.

We can use the results of section \ref{sec:3} to set constraints on the different leptoquark scenarios. This very much depends on the lepton generation the leptoquark couples to.

\begin{figure}[tbp]
\centering
\includegraphics[width=0.9\textwidth]{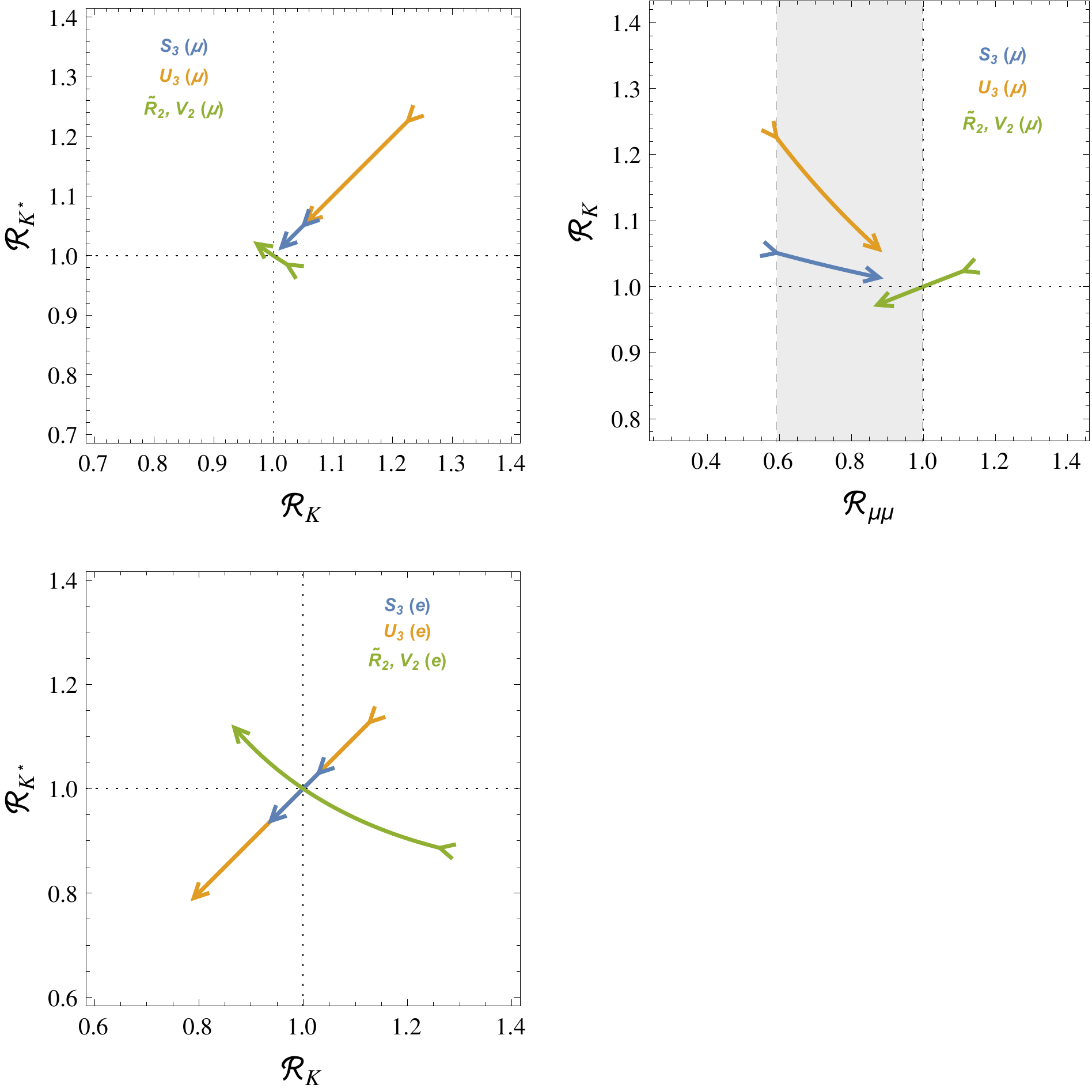}
\caption{Correlation between the branching fractions $B \rightarrow K \nu \bar \nu$, $B \rightarrow K^* \nu \bar \nu$ and $B_s \rightarrow \mu \mu$ for the different leptoquark models. We did not show the scenarios $S_1$ and $U_1$, since in the first case $\mathcal{R}_K = \mathcal{R}_{K^*}$ is in principle unbounded and in the second case we do not expect a deviation from the SM value. \emph{First line:} Only couplings to second generation leptons are allowed. \emph{Second line:} Only couplings to first generation leptons are allowed.}
\label{fig:Rplots-LQ}
\end{figure}

\paragraph{\boldmath$\ell=\mu$}
If we assume that the leptoquarks only couple to second generation leptons, all the bounds from $b \rightarrow s \mu^+ \mu^-$ decays apply, but they have to be rescaled by an appropriate factor depending on $n$. A slight complication arises for the scenario $V_2$, since for this case the quantum numbers of the leptoquark also allow for a second interaction term, such that additionally the operator $Q_{q e}$ is generated. This operator then carries completely independent Wilson coefficients, which do not contribute to the decay into neutrinos but potentially affect the bounds from the decay into charged leptons. Fortunately, this additional operator only contributes to the unprimed operators $C_9$ and $C_{10}$ such that the bounds on $\tilde c_{d l}$ and $\tilde c_{q e}$ are only weakly correlated. So, bounds from section \ref{sec:3} are only weakly modified.

The allowed values for the branching ratios are shown in figure \ref{fig:Rplots-LQ}.

\paragraph{\boldmath$\ell=e$}
If the leptoquarks only couple to first generation leptons then the bounds from section \ref{sec:3} apply. These are significantly weaker than for the case $\ell=\mu$. The resulting bounds on the branching ratios are also shown in figure \ref{fig:Rplots-LQ}. 

\paragraph{\boldmath$\ell=\tau$}
As already mentioned in section \ref{sec:3}, the bounds on the decays into taus are very weak. This means that effectively there are no bounds for this case.

\medskip
Until now, we assumed that the leptoquarks only couple to one generation of leptons. If we loosen this assumption we immediately get LFV. For the case of a coupling to third generation leptons we cannot find reasonable bounds due to the weak constraints from LFV processes involving taus. Thus, we only consider the case of non-vanishing couplings to the first two generations of leptons. For this we find that the following pattern of operators is generated schematically,
\begin{equation}
 \mathcal{L} \supset C_{s b ; e e} \left[  Q \right]_{s b;e e}+C_{s b ; \mu \mu} \left[  Q \right]_{s b;\mu \mu}+C_{s b ; e \mu} \left[  Q \right]_{s b;e \mu}+C_{s b ; \mu e} \left[  Q \right]_{s b;\mu e}\,\, + \,\, \mathrm{h.c.},
\end{equation}
where the Wilson coefficients obey the relation
\begin{equation}
 C_{s b ; e e } C_{s b; \mu \mu} = C_{s b ; e \mu} C_{s b ; \mu e}.
\end{equation}
We see that we can use the constraints on the flavour conserving Wilson coefficients to put constraints on the  flavour violating ones, which are only weakly bounded. Except for some fine-tuned corners of parameter space, we find that one cannot expect large effects
 in the flavour-violating Wilson coefficients, implying that charged LFV decays like $B_s\to e^\pm\mu^\mp$ are unlikely to be observable and that the contribution of the $b\to s\nu_e \bar\nu_\mu$ transition to the $B\to K^{(*)}\nu\bar\nu$ signal should be small.

Summarizing the results on leptoquarks, we can say that
 in the case of coupling to muons (and muon neutrinos), the effects in $\mathcal R_{K^{(*)}}$ can at most be of the order of 25\% for left-handed and 5\% for right-handed quarks. For couplings to electrons (and electron neutrinos), also right-handed currents can lead to deviations at the level of 20\% from the SM.

The exception to this is the scenario $S_1$ where the effects can in principle be very large, since this model is not constrained by $b \rightarrow s \ell^+\ell^-$ data at all. In all leptoquark models, if the leptoquark only couples to taus (and tau neutrinos), the effects in  $\mathcal R_{K^{(*)}}$ can be very large as well.

\section{Summary and conclusions}\label{sec:5}

In this paper, we have taken a close look at $b\to s\nu\bar\nu$ transitions, 
in view of the reduced form factor uncertainties, calculation of complete NLO electroweak corrections and new constraints from 
other decays, in particular $b\to s \mu^+\mu^-$ transitions. In addition to presenting improved SM predictions, we have 
analyzed these decays in  an effective field theory approach and in 
a number of explicit NP models. The numerous plots that exhibit the 
correlations between various observables demonstrate that these decays 
constitute a useful tool in constraining these models.

\begin{figure}[tbp]
\centering
\includegraphics[width=0.85\textwidth]{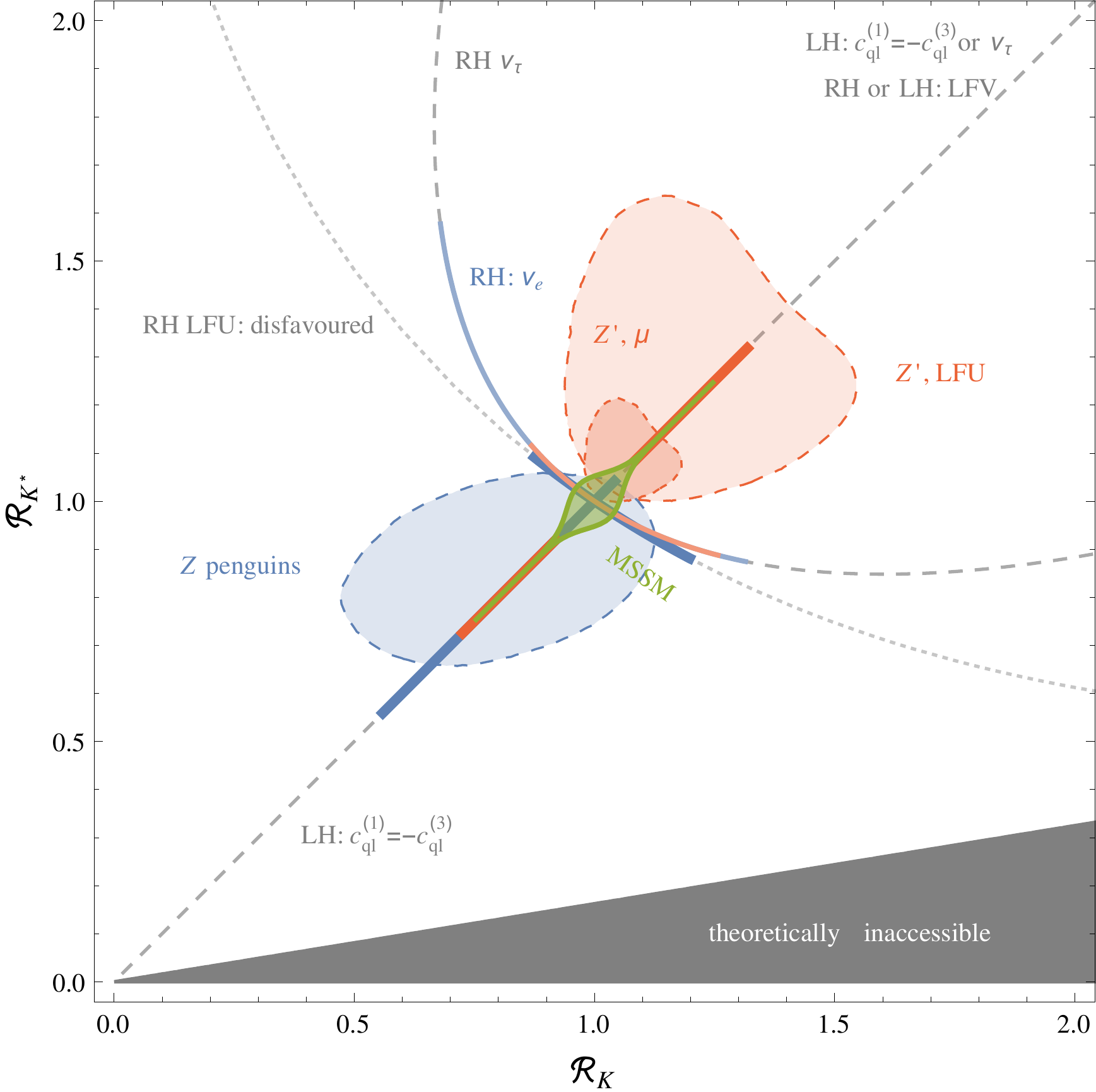}
\caption{Summary of allowed effects in the plane of $B\to K \nu\bar\nu$ vs. $B\to K^* \nu\bar\nu$ normalized to their SM values for various NP scenarios. For details see text.}
\label{fig:summary}
\end{figure}

Of particular interest 
is the correlation between $\mathcal{R}_K$ and $\mathcal{R}_K^{*}$.  In figure~\ref{fig:summary}, we collect the results for this 
correlation as obtained in several models:
\begin{itemize}
 \item The gray dashed line with $\mathcal{R}_K=\mathcal{R}_K^{*}$ is the MFV relation. The violation of this relation in future data  would  signal non-MFV interactions 
 and in particular right-handed currents at work. Staying on this line, there are three ways to obtain large deviations from the SM while avoiding $b\to s\ell^+\ell^-$ 
 constraints:
 \begin{itemize}
 \item Models generating the Wilson coefficients $\tilde c_{ql}^{(1)}=-\tilde c_{ql}^{(3)}$, e.g. the leptoquark model $S_1$ discussed in section~\ref{sec:lq}, can in principle lead to arbitrarily small or large effects;
 \item Models contributing only to operators with tau leptons and tau neutrinos can arbitrarily enhance the branching ratios, but there is a lower bound, $\mathcal{R}_K=\mathcal{R}_K^{*}>2/3$;
 \item Models contributing only to lepton flavour violating operators can only enhance the branching ratios. Interestingly, if NP only contributes to one LFV operator, the relation $\mathcal{R}_K=\mathcal{R}_K^{*}>1$ holds for left- and right-handed operators.
 \end{itemize}
 \item The gray dashed line labeled ``RH $\nu_\tau$'' is accessible in models contributing only to $(\tilde c_{dl})^{\tau\tau}$, i.e. right-handed currents with taus or tau neutrinos.
 \item The gray dotted line corresponds to models with right-handed currents only and with LFU. However, in the case of LFU, the $b\to s\ell^+\ell^-$ bounds apply, so large deviations from the SM are disfavoured.
 \item The coloured dark blue and red lines correspond to allowed effects when a single operator in the SM-EFT is varied. For details, see figure~\ref{fig:Rplots}.
 \item The light blue and red lines correspond to right-handed operators with electrons only. For details, see figure~\ref{fig:Rplots-ee}.
 \item The blue area corresponds to simultaneous effects in left- and right-handed $Z$ penguins (see figure~\ref{fig:BR-Z4f}).
 \item The large red area corresponds to  simultaneous effects in $\tilde c_{ql}^{(1)}$, $\tilde c_{dl}$, $\tilde c_{qe}$, and $\tilde c_{de}$ (see figure~\ref{fig:BR-Z4f}), assuming LFU, as happens in the presence of a single $SU(2)_L$ singlet $Z'$ gauge boson dominating the scene (cf.\ sec.~\ref{sec:genzp}).
 \item The small red area corresponds to the same operators as above, but assuming NP contributions only to operators with muons or muon neutrinos.
 \item The green area corresponds to the MSSM (see figure~\ref{fig:RK-SUSY}).
\end{itemize}

\bigskip\noindent
The main messages from this analysis are as follows:
\begin{itemize}
\item
The uncertainties in SM predictions for the branching ratios for $B\to K^{(*)} \nu\bar\nu$ have been reduced down to $10\%$.
\item
The SM branching ratio for $B\to K^*\nu\bar\nu$ is found to be by $40\%$ larger 
than previous estimates which could allow to observe this decay earlier than 
expected until now if we assume that NP contributions do not significantly suppress its rate.
\item
The precise measurements of decays based on the $b\to s \mu^+\mu^-$ transition can be used to put constraints on the size of effects in the $B\to K^{(*)} \nu\bar\nu$ decays, barring cancellations. Assuming LFU, this limits the relative deviations from the SM to roughly $\pm60\%$. If NP is assumed to affect only muons (and muon neutrinos), the effects are at most $\pm20\%$.
\item
We have emphasized that $b\to s\nu\bar\nu$ transitions could help to disentangle possible NP dynamics behind the anomalies presently observed in  $B\to K^{(*)} \mu^+\mu^-$ 
decays. This is seen in several plots presented by us, in particular in figures~\ref{fig:Rplots} and \ref{fig:BR-Z4f}.
\item
In the presence of flavour non-universality in the lepton couplings, NP effects in $b\to s \nu\bar\nu$ could be large, in particular if NP only couples to taus (and tau neutrinos).
In fact, the $B\to K^{(*)} \nu\bar\nu$ decays can be used to put indirect bounds on $b\to s\tau^+\tau^-$ transitions.
Some of the leptoquark models discussed by us represent concrete realizations of such a NP scenario.
This shows that,
without any dynamical 
assumptions, finding small NP effects in $b\to s \mu^+\mu^-$ transitions would 
not necessarily imply that in $b\to s \nu\bar\nu $ transitions these effects 
should also be small.
\item
In several NP scenarios, like the MSSM, 331 models, models with partial compositeness and models with MFV,
NP contributions to the branching ratios are not found larger than $30\%$ relative to their SM values.
\end{itemize}

In summary, our analysis demonstrates that the simultaneous study of the 
decays $B\to K^{(*)} \nu\bar\nu$, $B\to K^{(*)} \mu^+\mu^-$, and $B_s\to\mu^+\mu^-$ in the coming years will teach us a lot about the structure 
of possible new dynamics at the short distance scales in the reach of the LHC 
and even at much short distance scales. The measurement of the rate for 
$K^+\to\pi^+\nu\bar\nu$ by NA62 in  the coming years should also contribute in an important 
manner to these studies, possibly  signalling the presence of 
non-MFV interactions.

\section*{Acknowledgments}

It is a pleasure to thank Wolfgang Altmannshofer for useful discussions.
We are indebted to Aoife Bharucha and Roman Zwicky for sharing preliminary results on $B\to K^*$ form factors.
The research of A.B.\ and J.G.-N.\ was done and financed in the context of the ERC Advanced Grant
project ``FLAVOUR'' (267104). C.N.\ and D.S.\ were supported by the DFG cluster of excellence ``Origin and Structure of the Universe''.

\appendix

\section{Form factors}\label{sec:FF}\label{sec:kappa}

For the $B\to K^*$ form factors, we use the combined fit to lattice and LCSR results recently performed in \cite{BSZ}. 
The $\rho$ and $\kappa$ parameters used in section~\ref{sec:2} are defined in terms of $B\to K^*$ form factors as
\begin{align}
\rho_V(q^2)&=
\frac{2 q^2 \lambda^{3/2}_{K^*}(q^2)}{(m_B + {m}_{K^*})^2m_B^4}\left[V(q^2)\right]^2
\,,\\
\rho_{A_{1}}(q^2)&=
 \frac{2q^2\lambda^{1/2}_{K^*}(q^2)(m_B + {m}_{K^*})^2}{m_B^4}\left[A_1(q^2)\right]^2
\,,\\
\rho_{A_{12}}(q^2)&=
\frac{64   {m}_{K^*}^2 \lambda^{1/2}_{K^*}(q^2)}{m_B^2}  \left[A_{12}(q^2)\right]^2
\,,\\
\kappa_{\eta}(q^2) &= 2\,\frac{\rho_{A_{1}}(q^2)+\rho_{A_{12}}(q^2)-\rho_V(q^2)}{\rho_{A_{1}}(q^2)+\rho_{A_{12}}(q^2)+\rho_V(q^2)}
\,,
\end{align}
where
\begin{align}
\lambda(a,b,c)&= a^2  +b^2 + c^2 - 2 (ab+ bc + ac) \,,
&\lambda_{K^{(*)}}(q^2)\equiv\lambda(m_B^2,m_{K^{(*)}}^2,q^2)\,.
\label{eq:lambda}
\end{align}
In the case of binned observables, the correct definition of $\kappa_{\eta}$ to be used reads
\begin{align}
 \kappa_{\eta} |_{[a,b]} &=  2\,\frac{\int_a^b dq^2~(\rho_{A_{1}}(q^2)+\rho_{A_{12}}(q^2)-\rho_V(q^2))}{\int_a^b dq^2~(\rho_{A_{1}}(q^2)+\rho_{A_{12}}(q^2)+\rho_V(q^2))}
 \,.
\end{align}

For the $B\to K$ form factors, we proceed in a similar way.  We consider the $z$ expansion of the form factor $f_+$,
\begin{equation}
f_+(q^2) = \frac{1}{1-q^2/m_+^2}\left[
\alpha_0 + \alpha_1 z(q^2) + \alpha_2 z^2(q^2)+\frac{z^3(q^2)}{3}(-\alpha_1+2\alpha_2)
\right],
\end{equation}
where
\begin{equation}
z(t) = \frac{\sqrt{t_+-t}-\sqrt{t_+-t_0}}{\sqrt{t_+-t}+\sqrt{t_+-t_0}}\,,
\end{equation}
with
$t_\pm=(m_B\pm m_K)^2$ and $t_0=t_+(1-\sqrt{1-t_-/t_+})$.
The resonance mass is $m_+=m_B+0.046$\,GeV.
Results for the parameters $\alpha_0$, $\alpha_1$, and $\alpha_2$ have been presented in \cite{Bouchard:2013eph} based on a lattice computation valid at high $q^2$. Since we are interested in the full kinematical region, we add the additional information from LCSR that \cite{Ball:2004ye,Bartsch:2009qp}
\begin{equation}
f_+(0) = 0.304 \pm 0.042
\,,
\end{equation}
where we have used the updated value for the Gegenbauer moment in the $K^*$ distribution amplitude, $a_1^K=0.06 \pm 0.03$ \cite{Ball:2006fz,Buchalla:2008jp}. We then construct a $\chi^2$ function depending on the input from the lattice and the form factors (including $f_T$ and $f_0$) at $q^2=0$, retaining all known correlations. We determine the best-fit central values and (correlated) uncertainties of the $z$ expansion coefficients by marginalizing this $\chi^2$ function with a Markov Chain Monte Carlo algorithm. We find
\begin{align}
\alpha_0 &= 0.432 \pm 0.011
\,,&
\alpha_1 &= -0.664 \pm 0.096
\,,&
\alpha_2 &= -1.20 \pm 0.69
\,,
\end{align}
and the following correlation matrix,
\begin{equation}
\text{corr}(\alpha_i,\alpha_j) =
\begin{pmatrix}
1 & +0.32 & -0.37 \\
 +0.32 & 1 & +0.26 \\
 -0.37 & +0.26 & 1
\end{pmatrix}
.
\end{equation}
The parameter $\rho_K$ of section~\ref{sec:2} is defined as
\begin{equation}
\rho_{K}(q^2)=
\frac{\lambda^{3/2}_K(q^2)}{m_B^4}   \left[f^K_+(q^2)\right]^2,
\end{equation}
where $\lambda_K$ has been defined in (\ref{eq:lambda}).
\appendix

\bibliography{allrefs}
\bibliographystyle{JHEP}
\end{document}